\documentclass{aa}  
\usepackage{graphicx}
\usepackage{txfonts}
\usepackage{xcolor}

\usepackage{booktabs}
\usepackage{gensymb}
\usepackage{caption}

\usepackage[hyperindex, breaklinks=True, colorlinks, citecolor=blue, draft=False]{hyperref}

\authorrunning{Almeida et al.} 
\begin{document}

   \title{Forecasting synchrotron spectral parameters with QUIJOTE-MFI2 in combination with Planck and WMAP}

   \author{A. Almeida \inst{1,2} \fnmsep\thanks{E-mail: aalmeida@iac.es},
          J.~A. Rubiño-Martín \inst{1,2}, 
          R. Cepeda-Arroita \inst{1,2}, 
          R.~T. Génova-Santos \inst{1,2}
          \and 
          D. Adak \inst{1,2}
          }

   \institute{Instituto de Astrofísica de Canarias, Calle Vía Láctea S/N, E-38205 La Laguna, Tenerife, Spain 
         \and
             Universidad de La Laguna, Dpto. Astrofísica, Avda. Astrofísico Fco. Sánchez S/N, E-38206 La Laguna, Tenerife, Spain \\
             }

   \date{Received ; accepted }

    \abstract{We present a parametric component separation forecast for the QUIJOTE–MFI2 instrument (10--20\,GHz), assessing its impact on constraining polarised synchrotron emission at $1^\circ$ FWHM and $N_{\rm side}=64$. Using simulated sky maps based on power-law and curved synchrotron spectra, we show that adding QUIJOTE-MFI2 to existing WMAP+\textit{Planck}+MFI data yields statistically unbiased parameter estimates with substantial uncertainty reductions: improvement factors reach $\sim$10 for the synchrotron spectral index ($\beta_s$), $\sim$5 for the curvature parameter ($C_s$), and $\sim$43 for polarisation amplitudes in bright regions. Deep QUIJOTE cosmological fields enable $\beta_s$ constraints even in intrinsically low SNR regions where WMAP+\textit{Planck} alone remain prior-dominated. Current combined sensitivities are insufficient to detect a synchrotron curvature of $C_s=-0.052$ on a pixel-by-pixel basis, but a $2\sigma$ detection is achievable for $|C_s|\gtrsim 0.14$ in the brightest regions of the Galactic plane. In those deep cosmological fields, combining QUIJOTE-MFI2 with WMAP and \textit{Planck} reduces the median synchrotron residual at 100\,GHz by a factor of 6, to 0.033\,\textmu K$_{\rm CMB}$. These results demonstrate that QUIJOTE-MFI2 will provide critical low-frequency information for modelling Galactic synchrotron emission, offering valuable complementary constraints for future CMB surveys such as LiteBIRD and the Simons Observatory.
}

  \keywords{Cosmology: cosmic background radiation – Methods: data analysis}

   \maketitle

\section{Introduction} \label{sec:intro}

The detection of primordial B-modes is one of the most ambitious objectives of observational cosmology, as such a finding would provide direct evidence for inflation \citep[see, e.g.,][and references therein]{Bmodes_review}. However, this measurement faces several challenges, most notably contamination from polarised Galactic emission. Unlike the cosmic microwave background (CMB) intensity signal, the primordial B-mode signal is expected to be several orders of magnitude fainter than Galactic foregrounds across the entire sky \citep{2009AIPC.1141..222D, 2016A&A...594A..10P, 2020AA...641A...1P, 2020A&A...641A..11P}. At low frequencies (below $\sim$90\,GHz), polarised synchrotron emission, arising from relativistic electrons spiralling in the Galactic magnetic field, dominates on large angular scales. Alongside thermal dust emission at higher frequencies and instrumental systematics, it remains one of the main obstacles to detecting the faint primordial B-mode signal. Accurately measuring and subtracting synchrotron emission is therefore crucial to isolate the primordial B-mode signal from Galactic foregrounds.

For this reason, it becomes important to have a good spectral characterisation of the synchrotron emission. Achieving the resolution needed for B-mode extraction at a few GHz requires metre- to tens-of-metre-class dishes, making low-frequency telescopes too heavy for space deployment. Ground-based experiments such as S-PASS (2.3 GHz; \citealp{2019MNRAS.489.2330C}), C-BASS (5 GHz; \citealp{2018MNRAS.480.3224J}), and QUIJOTE (10–42 GHz; \citealp{mfiwidesurvey})\footnote{The combination of data from QUIJOTE, S-PASS, C-BASS, and \textit{Planck} to improve the characterisation of the polarised synchrotron emission is the goal of the Radioforegrounds+ project (\texttt{https://research.iac.es/proyecto/radioforegroundsplus/})} therefore play a crucial role in probing the synchrotron-dominated regime, where the signal is strongest. The high sensitivity of these experiments, coupled with the low-frequency coverage, fills a critical observational gap left by WMAP \citep{2013ApJS..208...20B} and \textit{Planck} \citep{2020AA...641A...1P}. Despite the advancements of WMAP and \textit{Planck} in the mapping of large scale polarised synchrotron, their lowest bands at 23 GHz and 28 GHz respectively, are limited in signal-to-noise ratio (SNR) especially at high Galactic latitudes \citep{2007ApJ...665..355K, 2020A&A...641A...4P}. As a consequence, the modelling of the synchrotron in these regions relies on simplified assumptions that neglect its spatial and spectral variability.

Synchrotron emission is usually modelled as a simple power law, though such a model neglects the complex spectral behaviour arising from Galactic structure and the physics of cosmic-ray electrons \citep{Longair2011synchrotron, 1986rpa..book.....R}. In reality, radiative losses of cosmic-ray electrons cause the synchrotron spectrum to steepen at higher frequencies (often referred to as spectral ageing), while the superposition of multiple synchrotron components along the line of sight and within the telescope beam can flatten or otherwise add higher-order deviations to the observed spectrum \citep{Davies1996, Bennett2003b, 2012ApJ...753..110K, 2017MNRAS.472.1195C}. In dense regions, synchrotron self-absorption can also occur, typically below $\nu\lesssim100$\,MHz. Modelling these effects is therefore essential to minimise foreground contamination, as neglecting them can introduce biases in the recovered CMB signal and undermine foreground-cleaning efforts \citep{2012MNRAS.424.1914A}.

Given the limitations of current data, only first-order departures from a power law are typically captured through a curvature term. But even measuring these first-order deviations is challenging, because low-frequency coverage is sparse, sensitivity is limited \citep{2012ApJ...753..110K, MFIcompsep_pol} and accurate relative calibration between datasets is required. 

Another difficulty arises in regions of low emission. In the faintest high-latitude areas, the synchrotron spectral index is poorly constrained by existing WMAP and \textit{Planck} data \citep{2020A&A...641A...4P, MFIcompsep_pol}. Such regions are of interest to cosmology, as their minimal foreground contamination provides the cleanest windows for probing the faint primordial B-mode signal. However, the weak constraints on the synchrotron spectral index in these areas limit the accuracy of foreground separation and increase the risk of residual contamination in the recovered CMB signal.

The combination of these challenges underscores the need for improved high-sensitivity, low-frequency observations. QUIJOTE-MFI2 (MFI2) is the upgraded multi-frequency successor to QUIJOTE’s first instrument, the Multi-Frequency Instrument (MFI). MFI provided polarimetric measurements in four channels at 11.1, 12.9, 16.8, and 18.8\,GHz, with angular resolutions of 0.9\textdegree - 0.6\textdegree, enabling the first detailed synchrotron spectral index map of the Northern Celestial Hemisphere at a 2$^{\circ}$ resolution when combined with WMAP and \textit{Planck} data \citep{mfiwidesurvey, MFIcompsep_pol, adak2025}. MFI2 \citep{2022SPIE12190E..33H} builds upon this with five polarimeter channels covering the same frequency range and 3 times greater raw sensitivity, confirmed through on-sky measurements, allowing a new wide survey with one year integration time to reach $15$\,\textmu K per $1^{\circ}$ beam at 11\,GHz - roughly equivalent to 1\,\textmu K arcmin at 100\,GHz when scaled with a power law assuming a synchrotron spectral index $\beta_s=-3.1$. The instrument is also equipped with a digital backend that samples the full frequency band, enabling the selective removal of sub-channels affected by Radio Frequency Interference from satellite constellations. 

In this work, we forecast the improvement in the estimation of polarised synchrotron emission enabled by the addition of the QUIJOTE-MFI2 channels to the existing WMAP, \textit{Planck} and QUIJOTE-MFI data. We examine the extent to which these data enhance the precision of $\beta_s$ measurements and whether the spectral index can be meaningfully constrained in low signal regions, where previous observations have provided only limited constraints. At the same time, we assess the prospects for detecting spectral curvature. Following a forecasting strategy similar for C-BASS \citep{2019MNRAS.490.2958J}, we use simulated sky maps and a parametric component separation method to compare foreground recovery with and without MFI2, considering different synchrotron models. 

This forecast is particularly relevant given the upcoming generation of CMB experiments. While thermal dust currently represents the dominant source of foreground uncertainty over most of the sky at CMB-relevant frequencies \citep{2016A&A...594A..10P, 2018A&A...618A.166K}, low-frequency synchrotron measurements, such as those provided by QUIJOTE-MFI2, will play an increasingly important role as dust characterisation from higher-frequency data becomes more accurate. This work therefore focuses on the synchrotron recovery and modelling performance of QUIJOTE-MFI2, providing constraints on the synchrotron spectral parameters that will inform Galactic foreground modelling in future CMB analyses, rather than targeting direct improvements to CMB parameter constraints at this stage. Looking ahead, as experiments like the Simons Observatory \citep{2019JCAP...02..056A} and LiteBIRD \citep{2023PTEP.2023d2F01L} achieve greater sensitivity, the synchrotron constraints provided by MFI2 will become valuable legacy information.

The structure of the paper is as follows. In Section \ref{sec:sim_data}, we describe the simulation of the polarised sky maps, including the instrumental specifications and the map processing treatment. Section \ref{sec:foregrounds} details the parametric foreground models, while Section \ref{sec:comp_separation} describes the component separation method used to fit these models to the data. In Sections \ref{sec:results_pixels} and \ref{sec:results_map} we present the results for both the QUIJOTE wide survey and the deep cosmology fields, beginning with four representative pixels and then extending the analysis to the full sky. Finally, Section \ref{sec:conclusions} summarises our conclusions.

\section{Simulated data} \label{sec:sim_data}

The forecasts presented in this work are based on simulations of the polarised microwave sky and instrumental response. In this section we describe the simulation of multi-frequency polarised sky maps, which provide the input to our component separation pipeline.

\subsection{Map generation} \label{sec:sim_data_maps}

Each frequency map is simulated using \texttt{PySM} \citep{2017MNRAS.469.2821T} at the centre frequency of each survey, at a \texttt{HEALPix} \citep{2005ApJ...622..759G} resolution of $N_{\text{side}} = 512$. The detailed characteristics of each survey, including frequency channels, beam sizes, and sensitivities are introduced in Section \ref{sec:sim_data_surveys}. The simulated sky consists of the CMB together with the dominant Galactic foregrounds, namely synchrotron and thermal dust emission. These components were modelled using the built-in \texttt{PySM} templates, adopting two different synchrotron spectral models (s1 and s3), while using the d1 and c1 models for dust and the CMB, respectively. Note that both s1 and s3 adopt the same spatially varying  spectral index map, but s3 additionally includes a curvature term assumed to be constant across the sky. 
A more detailed discussion of the different polarised foregrounds is presented in Section \ref{sec:foregrounds}.

To reproduce the observational response of each instrument, the simulated maps are convolved with a Gaussian beam of full-width at half maximum (FWHM) corresponding to the nominal beam size of the survey, $\theta_{\text{b}}$. Instrumental noise is added as independent realisations drawn from a Gaussian distribution with zero mean and standard deviation determined by the pixel-level sensitivity of each survey. We note that correlated noise, mainly instrumental and to a lesser extent atmospheric $1/f$ residuals in polarisation, is not simulated. Based on the existing MFI data \citep{mfiwidesurvey}, we expect the effective white noise on the pixel scale to only slightly increase due to the contribution of residual $1/f$ noise on the map after destriping. Each noise realisation is added to the corresponding frequency map, and the process is repeated one thousand times to build a robust statistical ensemble of simulated observations. All simulated maps are processed through the same preprocessing pipeline as usually applied to observational data. Each frequency map is smoothed to a common angular resolution of $1^{\circ}$ and subsequently downgraded to a \texttt{HEALPix} resolution of $N_{\text{side}} = 64$.

After smoothing and downgrading, the maps are multiplied by the survey masks appropriate to the QUIJOTE observations. For the wide survey forecasts we adopt the QUIJOTE default mask \citep{mfiwidesurvey}, which excludes areas contaminated by satellite signals, instrumental systematics near the North Celestial Pole, and noisy regions at low declinations. For the deep-field forecasts, we apply masks corresponding to the three QUIJOTE cosmological fields, which target low SNR regions optimised for cosmological analyses. In both cases, the mask is applied to all frequency maps, ensuring that forecasts across all surveys are consistently restricted to a common sky area (see Section \ref{sec:results_map} for a visualisation of the masked regions). A full description of the surveys is provided in the next section.

Finally, systematic effects such as colour correction, calibration uncertainties and beam errors are not included in this analysis. This forecast  therefore provides an initial assessment focused on the impact of instrumental sensitivity and angular resolution on the estimation of polarised synchrotron emission. The QUIJOTE instrument bandpasses are well characterised: the average colour corrections for the MFI bands are of the order of a few per cent, with uncertainties well below 1\%, and are thus expected to have a negligible impact on the inferred parameters. 

In the brightest regions of the sky, calibration uncertainty is expected to be the dominant systematic; for QUIJOTE, it is at the $\sim$5\% level, largely driven by uncertainties in the calibration source models \citep{mfiwidesurvey}. Since spectral curvature can only be constrained in these bright regions with the current sensitivity and sky coverage, unmodelled calibration uncertainties may lead to a modest underestimation of the errors on the recovered curvature parameter.

However, most of the remaining sky, corresponding to fainter, noise-dominated regions, is largely unaffected by these calibration errors. A detailed study of instrumental systematics is planned for future work.

\begin{figure*}
    \centering
    \includegraphics[width=\textwidth]{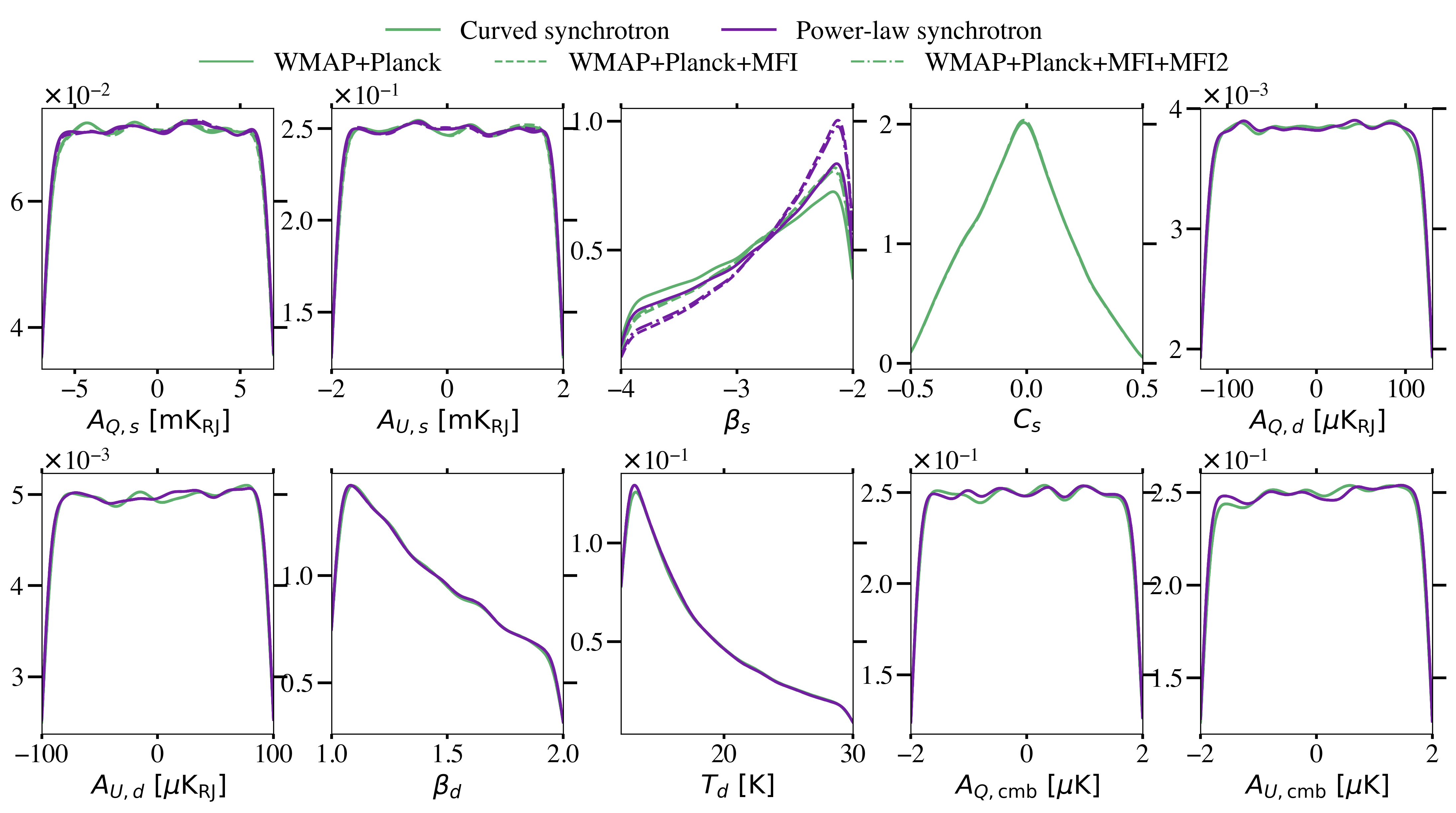}
    \caption{Independent Jeffreys prior distributions for each model parameter, computed for the three dataset combinations: WMAP+Planck, WMAP+Planck+MFI, and WMAP+Planck+MFI+MFI2. The curves show the two synchrotron models, with purple representing the power-law model (Eq. \ref{eq:model_sync_pl}) and green representing the curved model (Eq. \ref{eq:model_sync_curv}).}
    \label{fig:jeffreys_priors}
\end{figure*}

\subsection{Survey datasets and frequency coverage} \label{sec:sim_data_surveys}

This forecast focuses on the QUIJOTE-MFI2 instrument. In order to quantify the improvement in low-frequency polarisation measurements provided by MFI2, we compare its performance with publicly available datasets. Specifically, we consider WMAP \citep{2013ApJS..208...20B}, \textit{Planck} \citep{2020AA...641A...1P}, as well as the first QUIJOTE instrument, QUIJOTE-MFI \citep{mfiwidesurvey}. It is important to note that the sensitivity and frequency coverage of these higher-frequency datasets determine the extent to which MFI2 can improve the estimation of the synchrotron and CMB components.

The frequency coverage and characteristics of all the surveys are listed in Table \ref{tab:surveys} of Appendix \ref{sec:append_surveys}. For QUIJOTE-MFI and MFI2, we consider their two observing strategies: a wide area survey for Galactic characterisation and deep observations over cosmological fields for CMB studies. The QUIJOTE-MFI wide survey mapped approximately $29,000$ $\mathrm{deg}^2$ of the northern sky at 11–19 GHz, with one year of effective observing time, achieving polarisation sensitivities of 35–42 \textmu K per $1^{\circ}$ beam \citep{mfiwidesurvey}. Complementing the wide survey, three deep cosmological fields, each covering $\sim$ 1{,}000 $\mathrm{deg}^2$ and centred approximately at $(l,b) = (173.56^\circ, 47.09^\circ)$, $(78.93^\circ, 43.44^\circ)$, and $(124.39^\circ, -37.34^\circ)$, were observed with a total integration time of 2,200 hours per field. These regions were selected for their low Galactic foreground emission and are referred to as cosmological fields because of their suitability for CMB studies. The longer integration time in these fields reduces the noise per pixel relative to the wide survey, increasing the SNR for synchrotron measurements by about 1.7 for QUIJOTE-MFI and 2.5 for QUIJOTE-MFI2 (Table \ref{tab:surveys}).

The QUIJOTE-MFI2 will use the same observing modes as MFI. For MFI2, the instantaneous sensitivities are computed using the standard radiometer equation, assuming a system temperature of $T_{\text{sys}} = 17 \mathrm{K}$ \citep{2022SPIE12190E..33H}. The corresponding map sensitivities reported in Table~\ref{tab:surveys} are obtained by scaling these instantaneous sensitivities to account for half a year of effective observing time and the same sky coverage as the original MFI survey. 
Central frequencies and bandwidths for the new MFI2 maps are derived using a conservative approach, fully discarding the sub-bands between 10.7 and 12.7\,GHz, and 17.3--17.7\,GHz, which are affected by RFI contamination from downlinks of satellite megaconstellations \citep{2022SPIE12190E..33H}.
The calculation assumes a uniform distribution of observing time across the surveyed area. For the deep cosmological fields, the sensitivities are computed under the same assumption of half a year of effective observing time and are obtained by combining the MFI2 deep-field integrations with the wide survey data. Overall, MFI2 is expected to achieve polarisation sensitivities roughly 3--4 times better than QUIJOTE-MFI for the wide survey and 4--6 times better for the deep cosmological fields when combined with the wide survey data. However, it is important to note that the relevant metric for foreground characterisation is the effective SNR, rather than the absolute instrumental noise. For example, even though QUIJOTE-MFI2 has higher instrumental noise than \textit{Planck}, the synchrotron emission at 11–19 GHz is brighter, so in practice the effective SNR for synchrotron measurements can be roughly twice that of \textit{Planck}. Ultimately, both the effective SNR and frequency coverage define the overall constraining power.

\subsection{Estimation of true parameters and noise levels}

The beam convolution and resolution downgrading applied to the simulated maps described in Section \ref{sec:sim_data_maps} affect the effective theoretical data model. These processing steps modify the effective foreground spectral parameters at the resolution considered in the analysis, and also alter the pixel-level noise. As will be clear in Section \ref{sec:comp_separation}, properly accounting for these changes is essential. This characterisation of the effective “true” parameters and noise underpins the accurate measurement of biases and the correct estimation of errors.

Averaging the emission within the instrumental beam or over downgraded pixels mixes regions with different spectral properties, altering the effective parameters recovered at the final resolution. For example, averaging a pure synchrotron power-law over a beam or downgraded pixel no longer results in a perfect power law at the final resolution. This effect can be formally described using a moment expansion \citep{2017MNRAS.472.1195C}: if the spectral index or amplitude varies within a beam or pixel, the averaged signal differs from the value corresponding to the mean parameters, with the leading order term typically dominating when variations are relatively small.

As a result, the nominal input parameters cannot be assumed to represent the “true” values at the final resolution. We determine the effective "true" parameters empirically by performing pixel-by-pixel spectral fitting on the noiseless, processed maps using the L-BFGS-B least-squares method in \textsc{SciPy}. Throughout this work, these fitted parameters are treated as the true values of the parameters. To quantify the magnitude of these deviations in our specific simulation setup, we compute the SED in four representative sky regions (the Fan and Cygnus regions, the North Polar Spur and a low SNR area) and compare the spectral parameters recovered from the smoothed, downgraded maps against the \texttt{PySM} input values. We find deviations in $ \beta_s$ below 0.2 \% and in $C_s$ well below the forecasted uncertainties, meaning the processing steps introduce no significant biases in this case. This result is specific to the \texttt{PySM} s1 and s3 models, whose $\beta_s$ maps are already very smooth prior to any further processing (derived from the Haslam 408 MHz and WMAP 23 GHz data and smoothed with a 5$^\circ$ FWHM Gaussian), so that beam convolution and pixel downgrading mix only regions with nearly identical spectral indices. On the real sky, where spectral index variations occur on smaller angular scales and the true parameters are unknown, these deviations could be substantially larger and the empirical determination of the effective parameters described here would become essential for more realistic simulations. 

Consistent with this approach, the per-pixel noise levels are also affected by the processing. We estimate the noise as the standard deviation over one thousand independent noise realisations, each subjected to the same smoothing and downgrading described in Section \ref{sec:sim_data_maps}.

\section{Foreground modelling} \label{sec:foregrounds}

In this section, we present the parametric models used to fit simulated polarised Galactic foregrounds, summarising the spectral forms and assumptions of each component.

In polarisation, diffuse Galactic emission is dominated by synchrotron radiation at low frequencies and by thermal dust emission at higher frequencies, with a crossover around $\sim$ 70-90 GHz depending on the region of the sky \citep{2016A&A...594A..10P}. Secondary polarised components, though present in intensity, have a negligible contribution to the polarised sky. In particular, free–free emission, produced by electron-ion interactions in ionised gas, is intrinsically unpolarised. Minor polarisation can in principle be generated through Thomson scattering in anisotropic HII regions \citep{1986rpa..book.....R}, but this contribution is negligible on the diffuse sky. The free–free polarisation level is expected to be very low \citep{2016arXiv160603606D}, and observational constraints confirm the absence of significant signal, with upper limits at the level of a few percent ($\leq 3.4 $ \% at 95\% confidence; \citealt{2011MNRAS.418..888M}). Similarly, observations indicate that anomalous microwave emission (AME), whose mechanism is thought to be electric dipole emission from spinning dust grains, is effectively unpolarised, with upper limits on the polarisation fraction below 1–5 \% \citep{2011ApJ...729...25L, 2011MNRAS.418L..35D, W44, perseus_w43_rhooph}. Given their negligible polarisation, neither free–free nor AME are included in the polarisation forecasts presented in this work. Moreover, extragalactic radio sources can be polarised at the few percent level, but their contribution is negligible on the large angular scales of this forecast.

\begin{figure}
    \centering
    \includegraphics[width=\columnwidth]{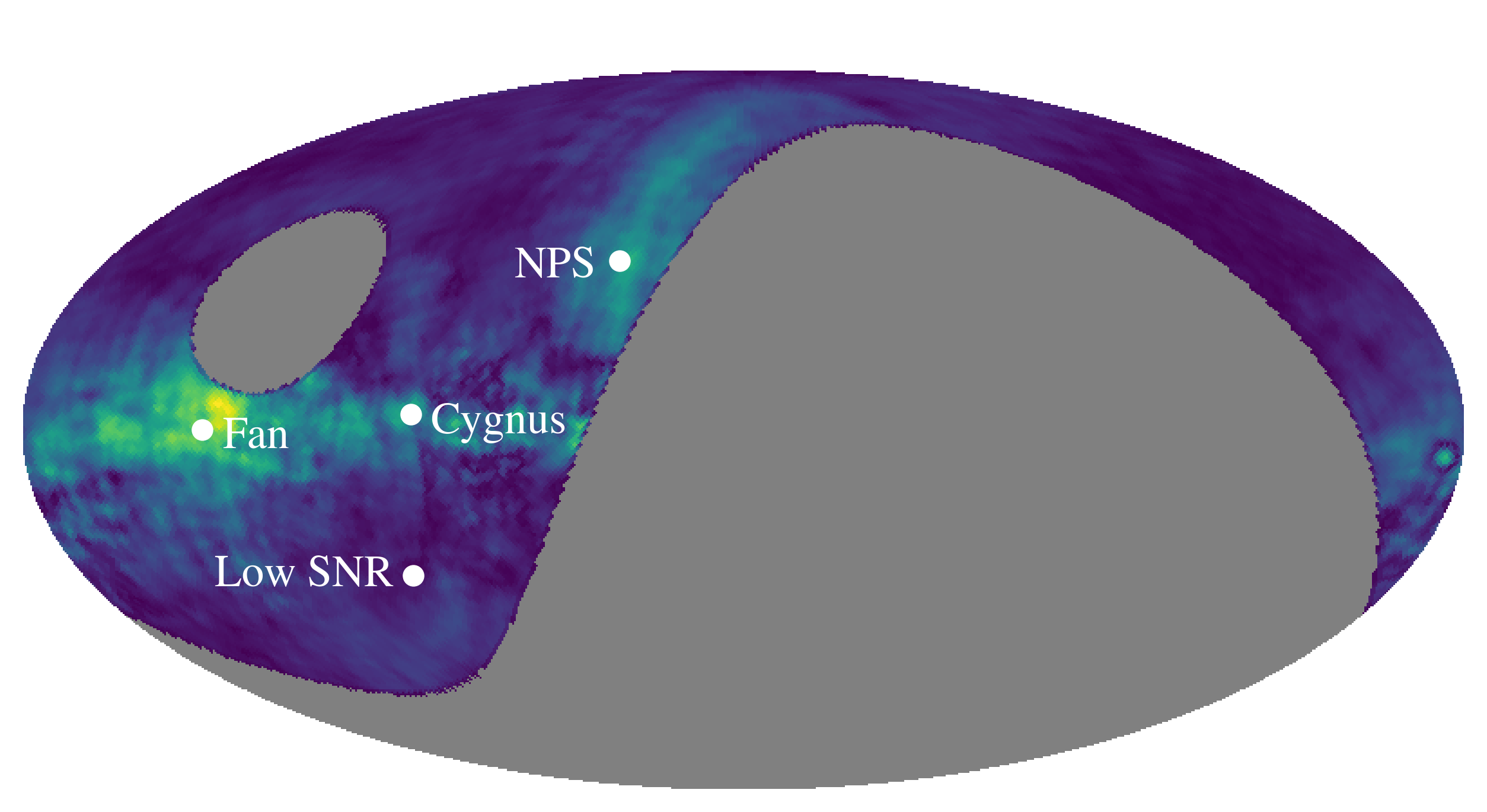}
    \caption{Location of the four case study pixels on the simulated 28.4 GHz polarisation intensity map.}
    \label{fig:pix_location}
\end{figure}

\begin{figure}
    \centering
    \includegraphics[width=\columnwidth]{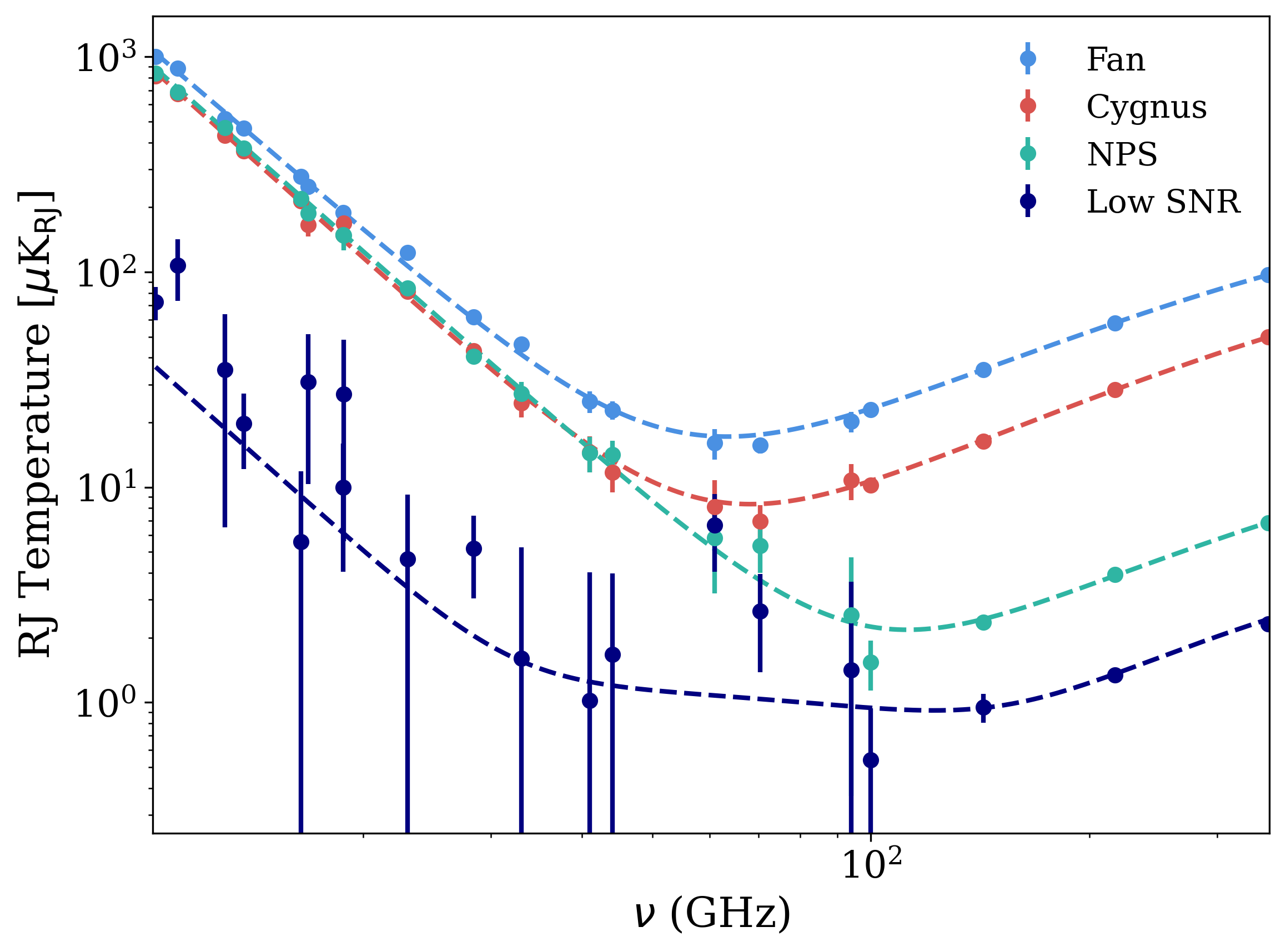}
    \caption{Spectral Energy Distribution (SED) of polarisation intensity $P = \sqrt{Q^2+U^2} $ of the four case study pixels. Dashed lines show the theoretical SED assuming a power law synchrotron model, while points represent simulated noisy data with corresponding error bars. No bias correction has been applied to $P$, which is plotted here for illustration only (all fits are performed on $Q$ and $U$ Stokes parameters). At low SNR the resulting Ricean bias\protect\footnotemark causes a slight deviation from the model, most evident for the low SNR pixel.}
    \label{fig:pix_sed}
\end{figure}

\footnotetext{The Ricean distribution describes the probability distribution of the magnitude of a vector whose components are independent Gaussian random variables \citep{rice1944mathematical}. For polarisation, the $Q$ and $U$ Stokes parameters form these components, so $P = \sqrt{Q^2 + U^2}$ follows a Rice distribution. Because $P$ is always positive, noise causes low SNR measurements to overestimate the true value, producing the Ricean bias.} 

We now discuss the parametric models used for the foreground separation in this work. These models follow the prescriptions from previous QUIJOTE papers \citep{MFIcompsep_pol, ameplanewidesurvey, perseus_w43_rhooph}. All models are expressed in Rayleigh-Jeans brightness temperature units.

\subsection{CMB} \label{sub_sec:cmb}
The CMB is polarised due to Thomson scattering of photons off free electrons in the surface of last scattering during recombination, and  later, during the reionisation era. CMB polarisation fluctuations can be expressed in Rayleigh-Jeans brightness temperature units as
\begin{equation}
\left[\begin{array}{c}
Q_{\mathrm{cmb}} \\
U_{\mathrm{cmb}}
\end{array}\right]=\left[\begin{array}{c}
A_{\mathrm{Q, cmb}}\\
A_{\mathrm{U, cmb}}
\end{array}\right]
\frac{x^2 \mathrm{e}^x}{\left(\mathrm{e}^x-1\right)^2},
\label{eq:model_cmb}
\end{equation}
where \( A_{\mathrm{Q, cmb}} \) and \( A_{\mathrm{U, cmb}} \) are the amplitudes in CMB brightness temperature for the Stokes parameters $Q$ and $U$, respectively, $ x \equiv (h \nu)/ ( k_{\mathrm{B}} T_o)$, with $\nu$ being the frequency, $h$ the \textit{Planck} constant, $k_{\mathrm{B}}$ the Boltzmann constant, and $T_o = 2.7255 $ K the mean CMB temperature \citep{2009ApJ...707..916F}.

\begin{figure*}
    \centering
    \includegraphics[width=\textwidth]{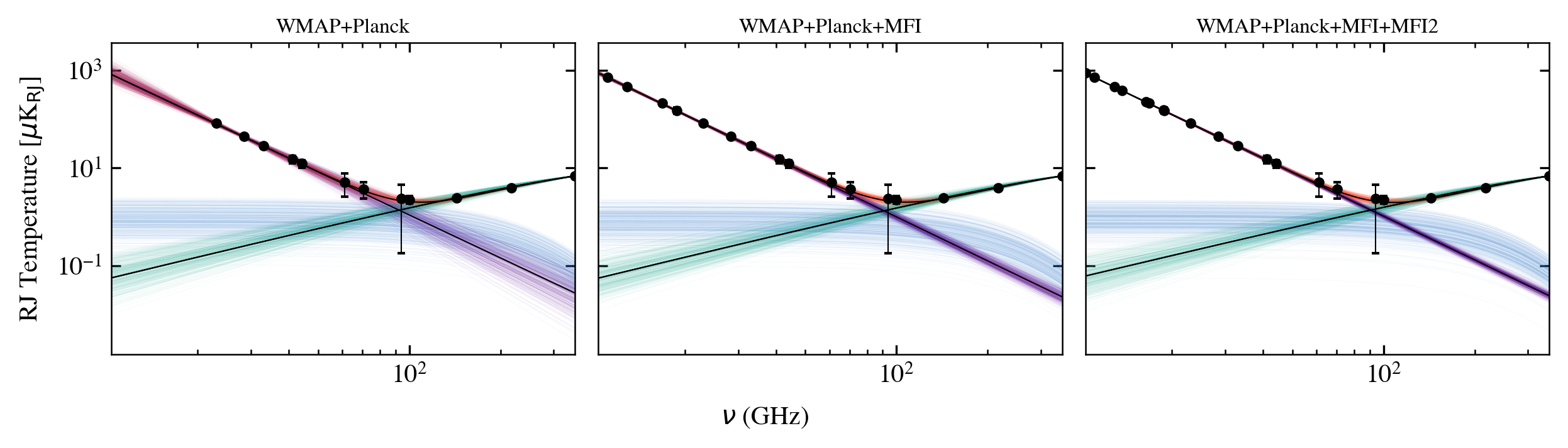}
    \caption{Polarisation intensity SED with MCMC samples for the NPS pixel. Colored lines show 400 random samples for the total model (red), synchrotron (purple), dust (turquoise), and CMB (blue) components. The black solid lines represent the best fit total SED and its individual components (synchrotron, dust, CMB). Data points with error bars correspond to the simulated observations.}
    \label{fig:sed_samples_nps}
\end{figure*}

\subsection{Synchrotron} \label{sub_sec:sync}

Synchrotron emission is produced by high-energy relativistic electrons moving in the Galactic magnetic field, so its spectrum depends on the electron energy distribution and the local field strength, both of which vary across the sky. The polarisation fraction can reach up to $\sim$75\% in regions with a uniform and ordered magnetic field \citep{2016arXiv160603606D}, but in practice it is reduced to 10–40\% in diffuse regions due to turbulence and line-of-sight depolarisation \citep{2014ApJ...790..104F, 2018A&A...618A.166K}. Faraday rotation, which rotates the direction of linear polarisation as the signal passes through magnetised plasma, is negligible at MFI2 frequencies ($\gtrsim$10 GHz) and does not significantly affect the observed polarisation intensity.

In the low-frequency regime, synchrotron emission is generally well described by a power-law spectrum \citep{1986rpa..book.....R}
\begin{equation}
\left[\begin{array}{c}
Q_s \\
U_s
\end{array}\right]=\left[\begin{array}{c}
A_{Q, s} \\
A_{U, s}
\end{array}\right]\left(\frac{\nu}{\nu_{\mathrm{0, s}}}\right)^{\beta_{\mathrm{s}}},
\label{eq:model_sync_pl}
\end{equation}
where \( A_{\mathrm{Q, s}} \) and \( A_{\mathrm{U, s}} \) are the $Q$ and $U$ synchrotron amplitudes at the pivot frequency $\nu_{\mathrm{0, s}}$ in brightness temperature units, and $\beta_s$ the spectral index, which is assumed to be spatially varying and identical for both $Q$ and $U$ Stokes parameters. While \texttt{PySM} \citep{2017MNRAS.469.2821T} adopts a pivot frequency of 23 GHz, we set the pivot to the lowest frequency in our data $\nu_{\mathrm{0, s}} = 10.35$ GHz (the rationale for this choice is discussed in the following section).

The power-law scaling is overly simplistic for describing the true frequency dependence of synchrotron emission. Even if the intrinsic emission follows a pure power law, averaging over a finite beam or pixel introduces deviations from a strict power-law shape. \citet{2017MNRAS.472.1195C} showed that such effects can be captured via a Taylor expansion of the spectrum, with the resulting moments describing the statistical properties of the effective spectral index. In practice, only a limited number of moments can be constrained, so the synchrotron spectrum is commonly approximated by a modified power law that allows the spectral index to vary logarithmically with frequency
\begin{equation}
\left[\begin{array}{c}
Q_s \\
U_s
\end{array}\right]=\left[\begin{array}{c}
A_{Q, s} \\
A_{U, s}
\end{array}\right]\left(\frac{\nu}{\nu_{0, s}}\right)^{\beta_s+C_s \log \left(\frac{\nu}{\nu_{0, s}}\right)}.
\label{eq:model_sync_curv}
\end{equation}
Here, the curvature parameter $C_s$ is related to the second moment of the spectral-index distribution, though Eq. (\ref{eq:model_sync_curv}) is not strictly a moment expansion. A negative $C_s$ produces spectral steepening at higher frequencies (positive $C_s$ flattens the spectrum). Physically, nonzero curvature can also arise from cosmic-ray electron ageing or from multiple synchrotron components along the line of sight. The s3 model of \texttt{PySM} \citep{2017MNRAS.469.2821T} adopts a small curvature value of $ C_{\mathrm{s}}  = -0.052 $, following the measurement of \cite{2012ApJ...753..110K}.

\subsection{Thermal dust} \label{sub_sec:dust}

Dust grains in the interstellar medium re-emit light as a modified blackbody, with polarisation arising from their alignment with the Galactic magnetic field. The simplest and most commonly used model assumes a single-temperature modified blackbody, characterised by a dust temperature $T_d$ and an emissivity spectral index $\beta_d$
\begin{equation}
\left[\begin{array}{c}
Q_{\mathrm{d}} \\
U_{\mathrm{d}}
\end{array}\right]=\left[\begin{array}{c}
A_{\mathrm{Q, d}} \\
A_{\mathrm{U, d}}
\end{array}\right]
\left(\frac{\nu}{\nu_{\mathrm{0, d}}}\right)^{\beta_{\mathrm{d}}+1} \frac{\exp(\gamma \nu_{\mathrm{0, d}} )-1}{\exp(\gamma \nu )-1},
\label{eq:model_dust}
\end{equation}
where \( A_{\mathrm{Q, d}} \) and \( A_{\mathrm{U, d}} \) are the thermal dust $Q$ and $U$ amplitudes at the pivot frequency $\nu_{\mathrm{0, d}}$, and $\gamma \equiv h / (k_B T_d) $.  
The pivot frequency is chosen to match \texttt{PySM} with $\nu_{0,d} = 353 $ GHz.

The total polarised emission is the sum of CMB, synchrotron and thermal dust emissions
\begin{equation}
\left[\begin{array}{c}
Q \\
U
\end{array}\right] =
\left[\begin{array}{c}
Q_{\mathrm{cmb}} \\
U_{\mathrm{cmb}}
\end{array}\right]
+
\left[\begin{array}{c}
Q_{\mathrm{s}} \\
U_{\mathrm{s}}
\end{array}\right]
+
\left[\begin{array}{c}
Q_{\mathrm{d}} \\
U_{\mathrm{d}}
\end{array}\right] .
\label{eq:model_total}
\end{equation}

\section{Component separation method} \label{sec:comp_separation}

In this work, we adopt a parametric component separation method to forecast the capability of QUIJOTE-MFI2 to disentangle the different sky components. Parametric component separation provides a controlled framework in which each emission mechanism is described by a parametric spectral model with some free parameters. This approach allows for the joint recovery of amplitudes and spectral parameters from simulated multifrequency data, providing direct physical interpretability and allowing instrumental noise and model uncertainties to be rigorously propagated into the  parameter constraints. Because this approach is model dependent, the accuracy of the recovered components is limited by the assumed models and incorrect modelling or incomplete frequency coverage can induce biases or leave residual foregrounds in the recovered components.

\begin{figure}
    \centering
    \includegraphics[width=\columnwidth]{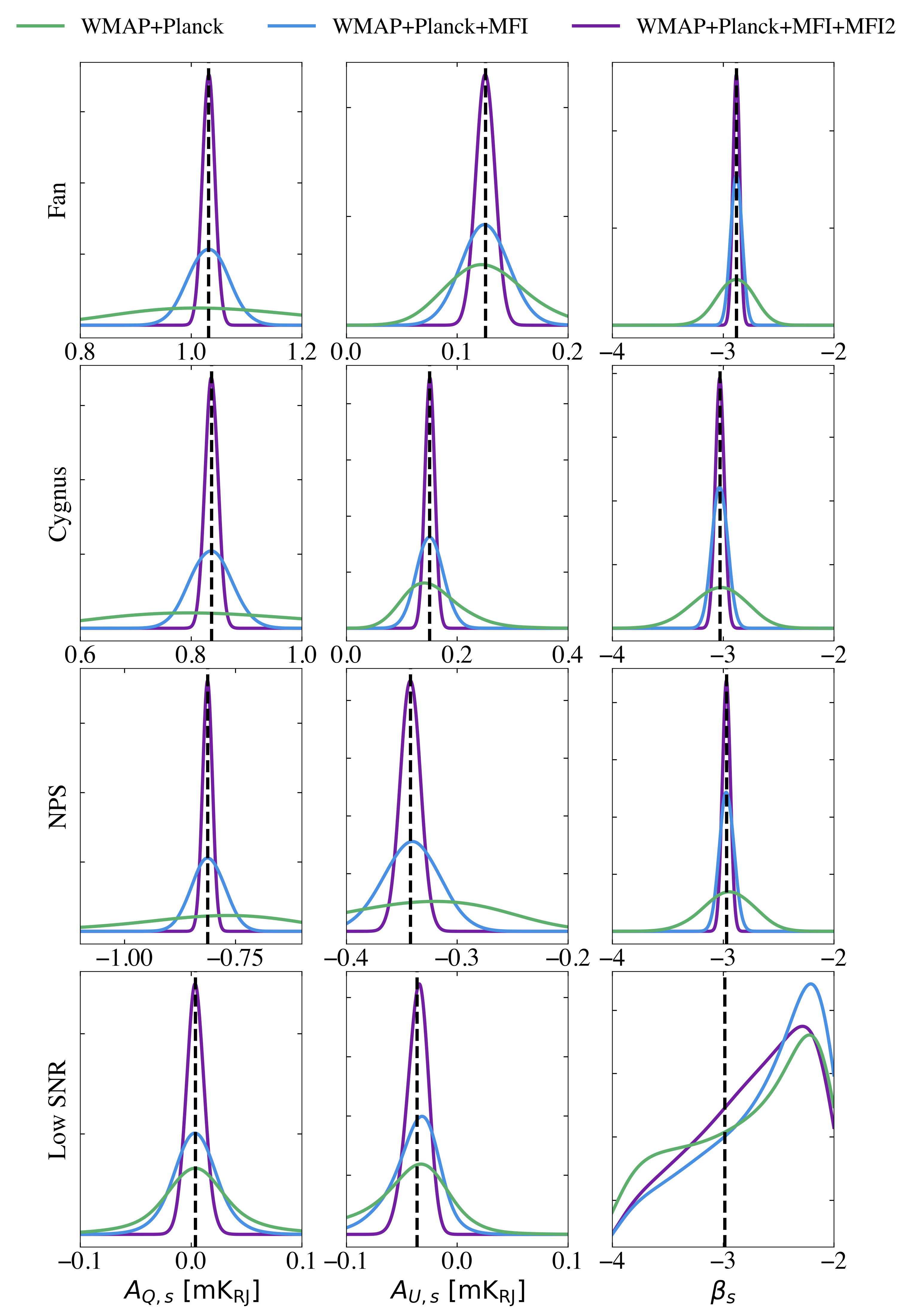}
    \caption{Marginalised one-dimensional posterior probability density functions (PDFs) for the synchrotron parameters in the four case study pixels in the wide survey configuration, assuming a synchrotron power-law spectrum. These were obtained from MCMC sampling with the three datasets: WMAP and \textit{Planck} in green, with the inclusion of MFI in light blue and with the inclusion of MFI2 in purple. The black dashed lines are the true parameter values known from \texttt{PySM}.}
    \label{fig:posteriors_no_curv_ws}
\end{figure}

\subsection{Bayesian inference} \label{sub_sec:bayesian}

In our analysis, parameter estimation is performed within a Bayesian framework on a pixel-by-pixel basis, allowing us to infer the full posterior distributions for all model parameters. The parametric models correspond to those introduced in Section \ref{sec:foregrounds}.

According to Bayes Theorem \citep{1763RSPT...53..269B}, the prior information about the $n$ free parameters $\boldsymbol{\theta} = (\theta_1, \theta_2, \dots, \theta_n)
$ of a model, is updated based on the data $\textbf{d}$, resulting in the posterior distribution
\begin{equation}
p(\boldsymbol\theta | \textbf{d}) \propto \pi(\boldsymbol\theta) \mathcal{L}(\textbf{d} |  \boldsymbol\theta),
\label{eq:bayes_theorem}
\end{equation}
where $\pi (\boldsymbol\theta)$ is the prior and $\mathcal{L}(\textbf{d} |  \boldsymbol\theta)$ the likelihood.

The noise in each frequency channel is modelled as Gaussian, so the brightness temperature measurements for the Stokes parameters $Q$ and $U$ at a given pixel are normally distributed around the true values $\bar{Q}_j$ and $\bar{U}_j$, with standard deviations given by the instrument sensitivity at that pixel, $\sigma_{Q,j}$ and $\sigma_{U,j}$. Assuming noise is uncorrelated between frequency channels, the joint likelihood for all $m$ channels is
\begin{equation}
\mathcal{L} \propto \exp \left[-\frac{1}{2} \sum_{j=1}^{m} \left(\frac{\left(Q_j-\bar{Q}_j\right)^2}{\sigma_{Q,j}^2} + \frac{\left(U_j-\bar{U}_j\right)^2}{\sigma_{U,j}^2}\right)\right],
\label{eq:likelihood}
\end{equation}
evaluated independently for each pixel in the map. In the case of QUIJOTE-MFI and MFI2, the 11 GHz and 13 GHz channels are correlated, and these correlations should be accounted for in the likelihood, as described in \cite{M31quijotemfi}. Neglecting such correlations between channels, as in the simplified likelihood above, still provides a reasonable approximation for forecasting purposes.

The posterior distribution is an n-dimensional surface of complex shape, which we explore using a Markov Chain Monte Carlo (MCMC) sampler \citep{2010CAMCS...5...65G}. In particular, we use the \texttt{emcee} implementation \citep{2013PASP..125..306F}, which utilises an affine-invariant ensemble algorithm that efficiently samples correlated parameters and scales well with high-dimensional spaces.

After testing various configurations, we find that running $50$ independent chains for $40,000$ steps, with an initial burn-in period discarding the first $20\%$ of samples, we achieve a good balance between computational time and convergence of the chains. We first ran the analysis for four representative pixels, during which the trace plots were visually inspected to assess convergence. We further monitor convergence using the autocorrelation time $\tau$, ensuring that the total number of steps satisfies $N_{\rm steps} \geq 50 \tau$. In a subsequent phase, convergence was evaluated for each pixel using the Geweke diagnostic \citep{geweke1992evaluating}.

We fit the Stokes parameters $Q$ and $U$ simultaneously. While this increases the dimensionality of the parameter space, it avoids the bias that arises when fitting the polarisation intensity $P = \sqrt{Q^2 + U^2}$, which follows a Rice distribution at low SNR due to its nonlinear dependence on $Q$ and $U$ \citep{2006PASP..118.1340V}. This approach is particularly important in low SNR pixels, where direct $P$ fitting would lead to less reliable parameter estimates.

\subsection{Choice of priors} \label{sub_sec:priors}

Bayesian inference depends on the choice of prior distributions (Eq. \ref{eq:bayes_theorem}), which encode prior knowledge about the model parameters. In the context of forecasts, the aim is to construct posteriors that depend solely on the data and the assumed models. For this reason, we opt for non-informative prior distributions.

Non-informative priors are designed to have a minimal influence on the inference. They are formal probability distributions defined such that the posterior is, by construction, dominated by the data. This means that their form is determined primarily by the structure of the likelihood. Jeffreys prior for example, is based on the Fisher information matrix which has information about the curvature of the likelihood, to create a prior that is invariant under reparameterisation \citep{1946RSPSA.186..453J}. Reference priors, on the other hand, are constructed by maximising the expected Kullback-Leibler divergence between the prior and posterior \citep{bernardo1979reference}.

When the parameter space is one dimensional, Jeffreys priors are invariant under reparameterisation because they scale with the square root of the Fisher information $\pi_J(\theta) \propto \sqrt{\mathcal{I} (\theta)}$, which transforms appropriately under a change of variable. When generalising this to higher dimensions, however, this invariance is not guaranteed to hold. The multivariate Jeffreys prior, defined through the determinant of the Fisher information matrix, can depend on the choice of parameterisation. Degeneracies between parameters and the marginalisation over nuisance parameters may result in the loss of invariance \citep{bernardo2009bayesian}.

\begin{table*}
    \centering
    \caption{Improvement factors on parameter constraints for the two QUIJOTE observation modes, wide survey and cosmological fields, for the two synchrotron models - synchrotron power law (PL) and curved synchrotron spectrum (C), defined as the ratio of the total errors of each parameter using WMAP+\textit{Planck} to those obtained with WMAP+\textit{Planck}+MFI+MFI2. Values greater than unity indicate improved constraints relative to WMAP and \textit{Planck}.}
    \label{tab:improv_f}
    \resizebox{\textwidth}{!}{
    \begin{tabular}{l*{20}{c}}
        \hline
         & \multicolumn{20}{c}{Improvement factors} \\
        \hline
        & \multicolumn{20}{c}{Wide survey} \\
        \hline
        Pixel & \multicolumn{2}{c}{$A_{Q,s}$} & \multicolumn{2}{c}{$A_{U,s}$} & \multicolumn{2}{c}{$\beta_s$} & \multicolumn{2}{c}{$C_s$} & \multicolumn{2}{c}{$A_{Q,d}$} & \multicolumn{2}{c}{$A_{U,d}$} & \multicolumn{2}{c}{$\beta_d$} & \multicolumn{2}{c}{$T_d$} & \multicolumn{2}{c}{$A_{Q,\mathrm{cmb}}$} & \multicolumn{2}{c}{$A_{U,\mathrm{cmb}}$} \\
        \cline{2-21}

         & PL & C & PL & C & PL & C & PL & C & PL & C & PL & C & PL & C & PL & C & PL & C & PL & C \\
        \hline
        Fan  & 16.1 &  42.4  & 3.5 & 6.1 & 5.3  & 9.8 & - & 4.5 & 1.0 & 2.5 & 1.0 & 1.2  & 1.0 & 1.2 & 1.1 & 1.1  & 1.0 & 1.0 & 1.1 & 1.2 \\

        Cygnus  & 19.5  & 32.6 & 5.1 & 6.9 & 6.0  & 7.6 & - & 3.3 & 1.0 & 1.7  & 1.0 & 1.0  & 1.0 & 1.0 & 1.0 & 0.9  & 1.1 & 1.0 & 1.0 & 1.2 \\

        NPS  & 17.8 & 33.0 & 7.8  & 14.0 & 6.3 & 8.4 & - & 3.7 & 1.0 & 2.4  & 1.0 & 1.5 & 1.0 & 1.0 & 1.0 & 1.0  & 1.4 & 1.9 & 1.2 & 1.2 \\

        Low SNR  & 3.4 & 3.7 & 4.1  & 4.4 & 1.3 & 1.2 & - & 1.3 & 1.0 & 1.0  & 1.0 & 1.0  & 1.0 & 1.0 & 1.0 & 1.0  & 1.1 & 1.1 & 1.0 & 1.0 \\
        
        \hline
        & \multicolumn{20}{c}{Cosmological fields} \\
        \hline
        Pixel & \multicolumn{2}{c}{$A_{Q,s}$} & \multicolumn{2}{c}{$A_{U,s}$} & \multicolumn{2}{c}{$\beta_s$} & \multicolumn{2}{c}{$C_s$} & \multicolumn{2}{c}{$A_{Q,d}$} & \multicolumn{2}{c}{$A_{U,d}$} & \multicolumn{2}{c}{$\beta_d$} & \multicolumn{2}{c}{$T_d$} & \multicolumn{2}{c}{$A_{Q,\mathrm{cmb}}$} & \multicolumn{2}{c}{$A_{U,\mathrm{cmb}}$} \\
        \cline{2-21}
         & PL & C & PL & C & PL & C & PL & C & PL & C & PL & C & PL & C & PL & C & PL & C & PL & C \\
        \hline    
        Low SNR  & 10.1 & 11.0 & 10.2 & 11.0  & 1.8 & 1.6 & - & 1.2  & 1.0 & 1.0  & 1.0 & 1.0  & 1.0 & 1.0 & 1.0 & 1.0  & 1.1 & 1.1 & 1.1 & 1.1 \\
        \hline 

    \end{tabular}
    }
\end{table*}

To avoid the parameterisation issues associated with the multivariate Jeffreys prior, we adopt the independent Jeffreys prior \citep{2008ApJ...676...10E, bernardo2009bayesian}, which treats the parameters as independent. For a parameter $\theta_i$, the prior is defined as
\begin{equation}
\pi_{J}(\theta_i)  \propto \sqrt{\det \mathcal{I}_{ii}(\theta_i)},
\label{eq:jeffreys}
\end{equation}
where $ \mathcal{I}_{ii}$ denotes the $i$-th diagonal element of the Fisher information matrix:
\begin{equation}
\mathcal{I}_{ii}(\theta_i) = - \mathbb{E} \left[\frac{\partial^2 \ln \mathcal{L}}{\partial \theta_i^2} \right].
\label{eq:fisher}
\end{equation}

Assuming the parameters are a priori uncorrelated, the full prior is constructed as the product of the one-dimensional Jeffreys priors,
\begin{equation}
\pi(\boldsymbol{\theta}) \propto \prod_i \pi_{\mathrm{J}} \left(\theta_i\right)
\label{eq:full_prior}.
\end{equation}

The analytical expressions of the Independent Jeffreys prior for the parameter space of this work are given in Table \ref{tab:jeffreys_priors} of Appendix \ref{sec:append_jeffreys} and their corresponding distributions are plotted in Figure \ref{fig:jeffreys_priors}. The upper and lower bounds placed on the priors are informed by the minimum and maximum values of the theoretical parameter maps from \texttt{PySM}. 

An important nuance of the independent Jeffreys prior is that it depends not only on the model but also on the dataset, because the Fisher information incorporates both the model parameters and data characteristics such as noise levels and frequency coverage. For example, as seen in Table \ref{tab:jeffreys_priors}, the prior for the synchrotron spectral index $\beta_s$, depends on the choice of pivot frequency, $\nu_{\mathrm{0,s}}$ and its behaviour varies depending on how the dataset’s frequency coverage relates to this pivot. For instance, setting $\nu_{\mathrm{0,s}} = 23$ GHz produces a prior that differs between low-frequency datasets (e.g., QUIJOTE-MFI2) and higher-frequency datasets (e.g., WMAP and \textit{Planck}). Although mathematically correct, this introduces inconsistencies when comparing datasets. To avoid this, we adopted $\nu_{\mathrm{0,s}} = 10.35$ GHz, which corresponds to the lowest frequency in our combined datasets. This ensures that the Jeffreys prior is consistent across both low-frequency (QUIJOTE) and high-frequency (WMAP, Planck) data.

In addition to the Jeffreys prior, which we adopt in the case-study pixel analysis (Section \ref{sec:results_pixels}), we also consider Gaussian priors for the map-level forecasts (Section \ref{sec:results_map}). While Jeffreys priors are well suited for high SNR regions, they become ineffective in the low SNR regime that dominates the majority of the sky, where informative constraints are needed for reliable synchrotron recovery. To enable a direct comparison with previous component separation studies based on real QUIJOTE-MFI data \citep{MFIcompsep_pol}, we adopt the same Gaussian priors as in that work.

Moreover, to prevent the parameters from taking unphysical values, we impose a constraint on the effective synchrotron spectral index, following the approach of \cite{2019MNRAS.490.2958J}
\begin{equation}
-4 \leq \beta_{eff} \equiv \beta_s + C_s \log\bigg(\frac{500}{v_{0,s}}\bigg) \leq -2.
\label{eq:beta_eff}
\end{equation}

\subsection{Metrics of improvement} \label{sub_sec:metrics_improv}

We evaluate the impact of QUIJOTE-MFI2 on synchrotron parameter estimation by comparing posterior uncertainties, biases, and overall accuracy with and without its inclusion.

An estimator $\hat{\theta}_i$ has an associated standard error, given by the standard deviation
\begin{equation}
\operatorname{var}(\hat{\theta}_i) = \mathbb{E}\left[(\hat{\theta}_i - \mathbb{E}[\hat{\theta}_i])^2\right],
\label{eq:var}
\end{equation}
where $\mathbb{E}[\cdot]$ denotes the ensemble average, computed over 1,000 realisations. The estimation error arises not only from the statistical uncertainty but also from systematic deviations, quantified by the bias. The bias measures the difference between the expected value of the estimator and the true parameter value, and is defined as
\begin{equation}
\operatorname{bias}(\hat{\theta}_i) = \mathbb{E}[\hat{\theta}_i] - \theta_i.
\label{eq:bias}
\end{equation}
The estimator is unbiased if its expected value equals the true parameter, $\mathbb{E}[\hat{\theta}_i] = \theta_i$. The mean squared error (MSE) of an estimator $\hat{\theta}_i$ combines both variance and bias into a single metric:
\begin{equation}
\mathrm{MSE}(\hat{\theta}_i)= \mathbb{E}\left[(\hat{\theta}_i - \theta_i)^2\right] = \operatorname{var}(\hat{\theta}_i) + \operatorname{bias}^2(\hat{\theta}_i).
\label{eq:mse}
\end{equation}

The improvement factors are then defined as the ratio of the root mean squared errors (RMSE) estimated using one dataset to those obtained with another
\begin{equation}
\mathrm{RMSE}(\hat{\theta}_i) = \sqrt{\mathrm{MSE}(\hat{\theta}_i)} = \sqrt{\mathbb{E}\left[(\hat{\theta}_i - \theta_i)^2\right] }.
\label{eq:rmse}
\end{equation}

\section{Results: case study pixels} \label{sec:results_pixels}

We begin by analysing a set of representative case study pixels, before extending the analysis to the full maps in Section \ref{sec:results_map}. We select four pixels from representative regions with different relative contributions of synchrotron and thermal dust emissions. Specifically, we choose one pixel each in the Fan, Cygnus, and North Polar Spur (NPS) regions, and a fourth in a low SNR area. Their locations on the sky are shown in Figure \ref{fig:pix_location}, and their corresponding spectral energy distributions (SEDs) are plotted in Figure \ref{fig:pix_sed}.

\begin{figure}
    \centering
    \includegraphics[width=\columnwidth]{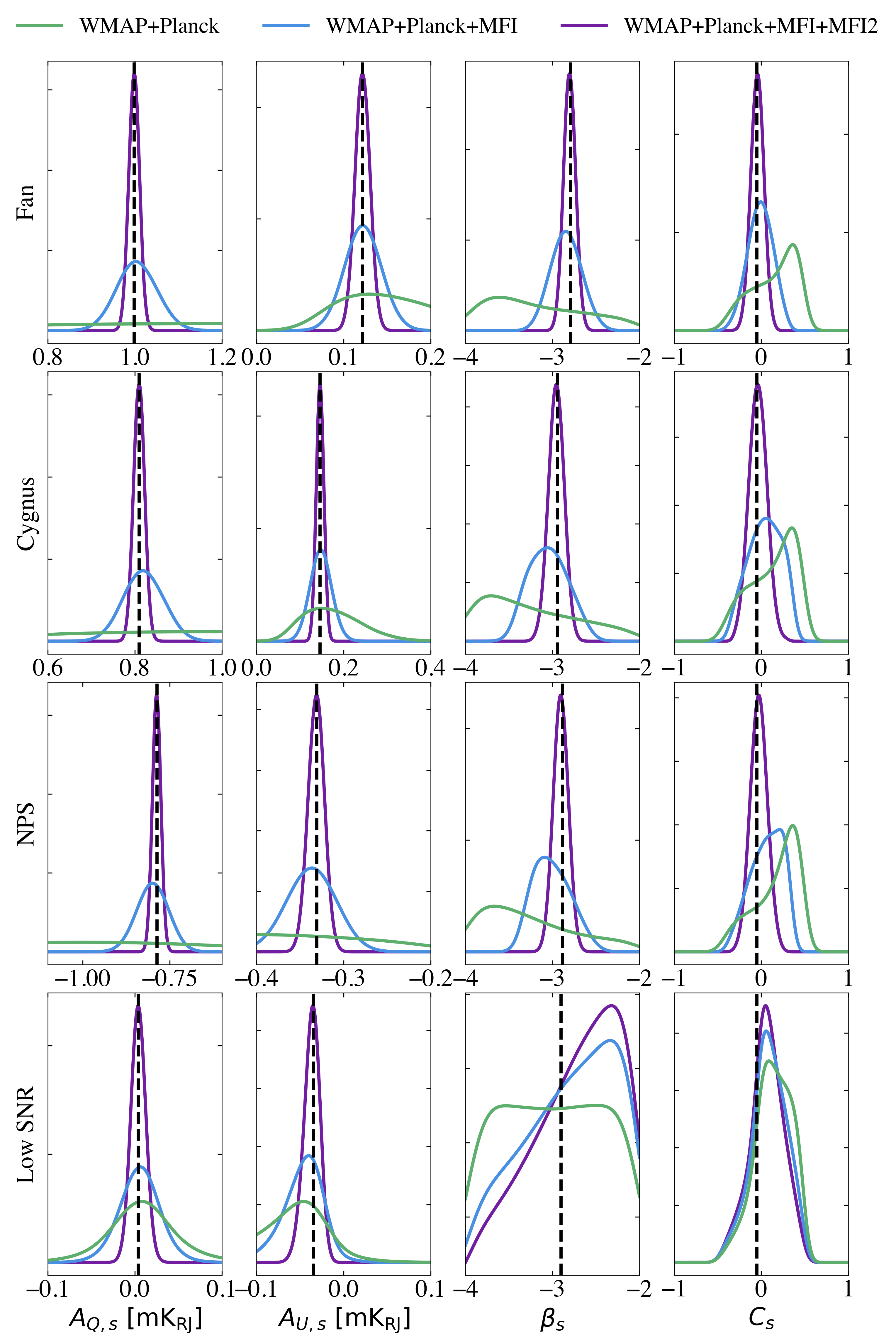}
    \caption{1D marginalised PDFs for the synchrotron parameters in the four case study pixels in the wide survey configuration, assuming curved synchrotron spectrum, for the three datasets. The black dashed lines are the true parameter values known from \texttt{PySM}.}
    \label{fig:posteriors_curv_ws}
\end{figure}

The forecasts target MFI2, but a baseline is required for comparison. For this, we consider existing public polarisation data: WMAP and \textit{Planck}, as well as the predecessor QUIJOTE instrument, MFI. The characteristics of all these datasets were summarised in Table \ref{tab:surveys}. Forecasts are then performed for the three datasets, namely WMAP+\textit{Planck}, WMAP+\textit{Planck}+MFI, WMAP+\textit{Planck}+MFI+MFI2, allowing us to quantify the improvement brought by MFI2 over previous measurements.

The pixel–level forecasts are carried out in two ways. First, we use the QUIJOTE wide survey sensitivities to forecast all four representative pixels. Second, we consider the QUIJOTE cosmological fields sensitivities (combined with the wide survey ones), to determine the improvement achievable in the low SNR pixel. These forecasts are performed for two sets of simulated data, which are also fit with the corresponding model: (i) a simple power–law synchrotron (Eq. \ref{eq:model_sync_pl}), and (ii) a curved synchrotron spectrum (Eq. \ref{eq:model_sync_curv}).

As a final remark, we note that our analysis is primarily aimed at assessing the improvement on the characterisation of the polarised synchrotron emission. Given the data considered here, where the high-frequency Planck channels set the dominant uncertainties on thermal dust, adding the low-frequency QUIJOTE channels has, as expected, little impact on dust parameter recovery.

\subsection{Wide survey results}  \label{sec:results_pixels_ws}

\subsubsection{Power-law synchrotron} \label{sec:results_pixels_ws_pl}

The first set of forecasts addresses the simplest scenario, with synchrotron emission following a power–law spectrum (Eq. \ref{eq:model_sync_pl}), under the wide survey sensitivities. Figure \ref{fig:sed_samples_nps} presents 400 random samples from the MCMC chains, showing that the inclusion of the MFI2 wide survey data points reduces the scatter of both the synchrotron component and the overall fit.

Figure \ref{fig:posteriors_no_curv_ws} shows the marginalised posterior probability distributions (PDFs) of the parameters for the four case-study pixels, plotted for the noiseless case. In practice, the noiseless posterior can be approximated by computing the posterior for each of many independent noise realisations and then taking the ensemble average. For a large number of realisations, random noise contributions largely cancel, and the ensemble average closely approaches the noiseless posterior. This approximation is essentially exact at high SNR, but at low SNR the posterior depends nonlinearly on the data, so the average provides only an approximate representation of the true noiseless distribution. Table \ref{tab:improv_f} shows the improvement factors in parameter estimation achieved by adding QUIJOTE data, relative to the constraints obtained with WMAP and \textit{Planck} alone.

For the high SNR pixels (Fan, Cygnus, NPS), the forecasts show that including the low-frequency QUIJOTE data leads to a clear improvement in the estimation of the synchrotron spectral parameters. In the case of WMAP+\textit{Planck} alone, the uncertainties for the spectral index $\beta_s$ are substantial, in the range of 0.16-0.23. MFI reduces these uncertainties to 0.05-0.07, while the inclusion of MFI2 further tightens the constraints to 0.03-0.04. This translates into improvement factors over WMAP+\textit{Planck} of approximately 5–6 with MFI2. Naturally, the ability to constrain the synchrotron parameters depends on the intrinsic synchrotron brightness of that pixels, with brighter pixels such as the Fan one, yielding tighter constraints.

A similar trend is seen for the synchrotron amplitudes. Adding MFI2 reduces the total errors relative to WMAP+\textit{Planck} by factors of 4-5 for $A_{Q,s}$ and 1.7-3.5 for $A_{U,s}$, yielding improvement factors of $\sim$ 16-20 for $A_{Q,s}$ and $\sim$4-8 for $A_{U,s}$, depending on the pixel. We note, however, that the amplitude uncertainties are partly limited by the choice of pivot frequency $\nu_{\mathrm{0,s}} = 10.35 $ GHz, which lies below the lowest frequencies covered by WMAP and \textit{Planck}, naturally inflating their uncertainties at this reference frequency. As illustrated by the synchrotron amplitude PDFs in Fig. \ref{fig:posteriors_no_curv_ws}, their ability to constrain the amplitude at $\nu_{\mathrm{0,s}}$ is limited. We do not discuss parameter biases here, as they are negligible for a pure power–law model. The improvement factors therefore reflect primarily the reduction in statistical uncertainties.

\begin{table}
    \centering
    \captionsetup{font=small}
    \begin{tabular}{c r r r r}
        \toprule
        \multicolumn{5}{c}{Normalised bias} \\
        \midrule
        Pixel & $A_{Q,s}$ & $A_{U,s}$ & $\beta_s$ & $C_s$ \\
        \midrule
        \multicolumn{5}{c}{WMAP+\textit{Planck}+MFI+MFI2} \\
        \cmidrule(lr){1-5} 
        Fan       &  0.05 &  0.01 & -0.10 &  0.13 \\
        Cygnus    &  0.09 &  0.05 & -0.12 &  0.12 \\
        NPS       & -0.11 & -0.10 & -0.16 &  0.22 \\
        Low SNR   &  0.08 & -0.22 &  0.41 &  0.68 \\
        \midrule
        \multicolumn{5}{c}{WMAP+\textit{Planck}} \\
        \cmidrule(lr){1-5} 
        Fan       &  0.66 &  0.47 & -0.75 &  0.84 \\
        Cygnus    &  0.53 &  0.37 & -0.67 &  0.78 \\
        NPS       & -0.60 & -0.57 & -0.77 &  0.96 \\
        Low SNR   &  0.11 & -0.29 & -0.13 &  0.92 \\
        \bottomrule
    \end{tabular}
    \caption{Normalised bias values of the synchrotron spectral parameters (i.e., ratio of the deviations relative to their uncertainties) for the four case-study pixels, for the wide survey mode under the curved synchrotron model, for WMAP+\textit{Planck} and WMAP+\textit{Planck}+MFI+MFI2 datasets.}
    \label{tab:bias_curv}
\end{table}

\begin{figure}
    \centering
    \includegraphics[width=\columnwidth]{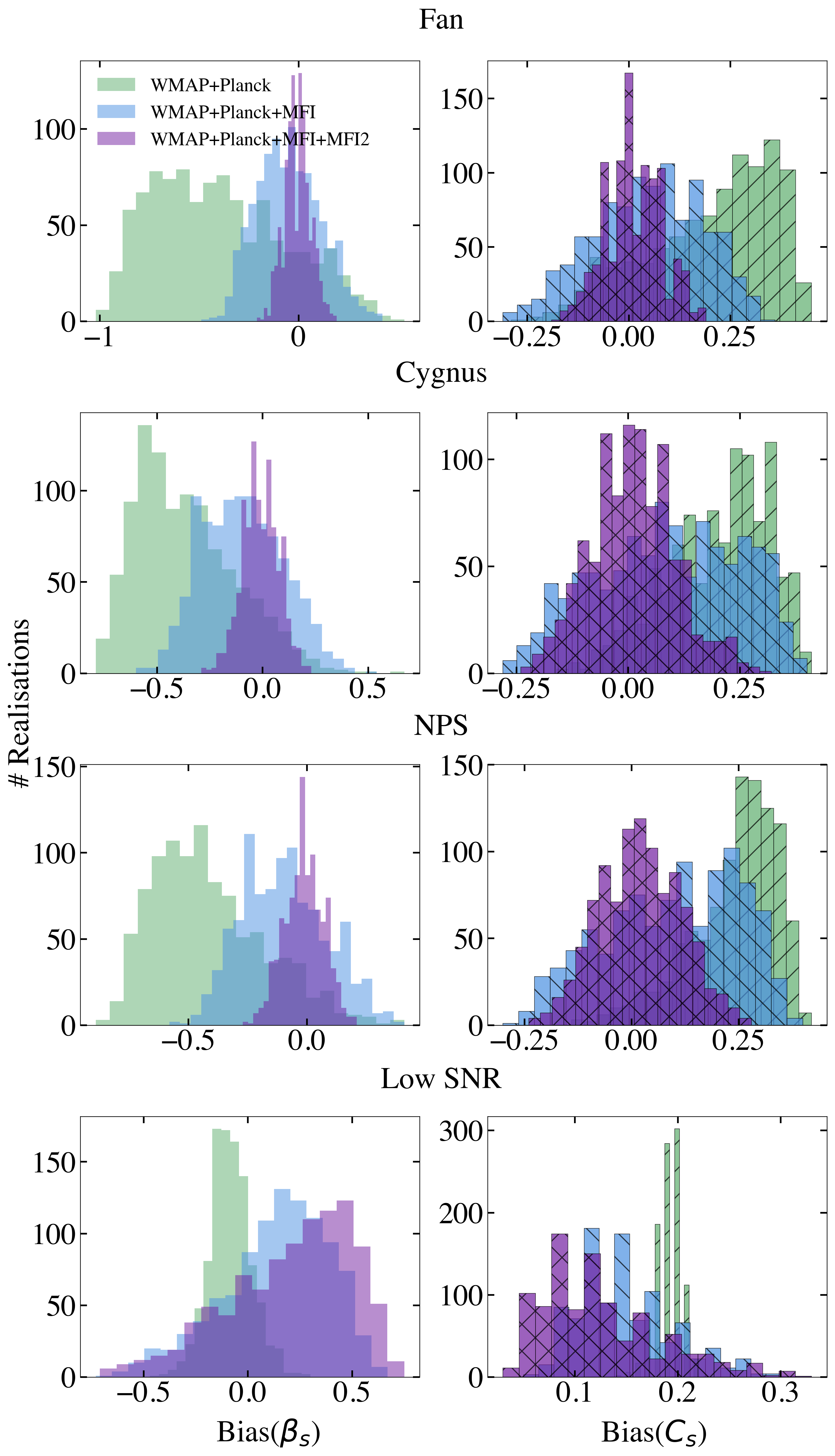}
    \caption{Forecast bias distribution of the synchrotron spectral parameters for a curved synchrotron spectrum with wide survey sensitivities, based on a thousand realisations. Each row corresponds to one of the four case-study pixels. Left column: bias in $\beta_s$; right column: bias in $C_s$}
    \label{fig:bias_hist_BsCs_curv}
\end{figure}

\begin{figure}
\centering
\includegraphics[width=\columnwidth]{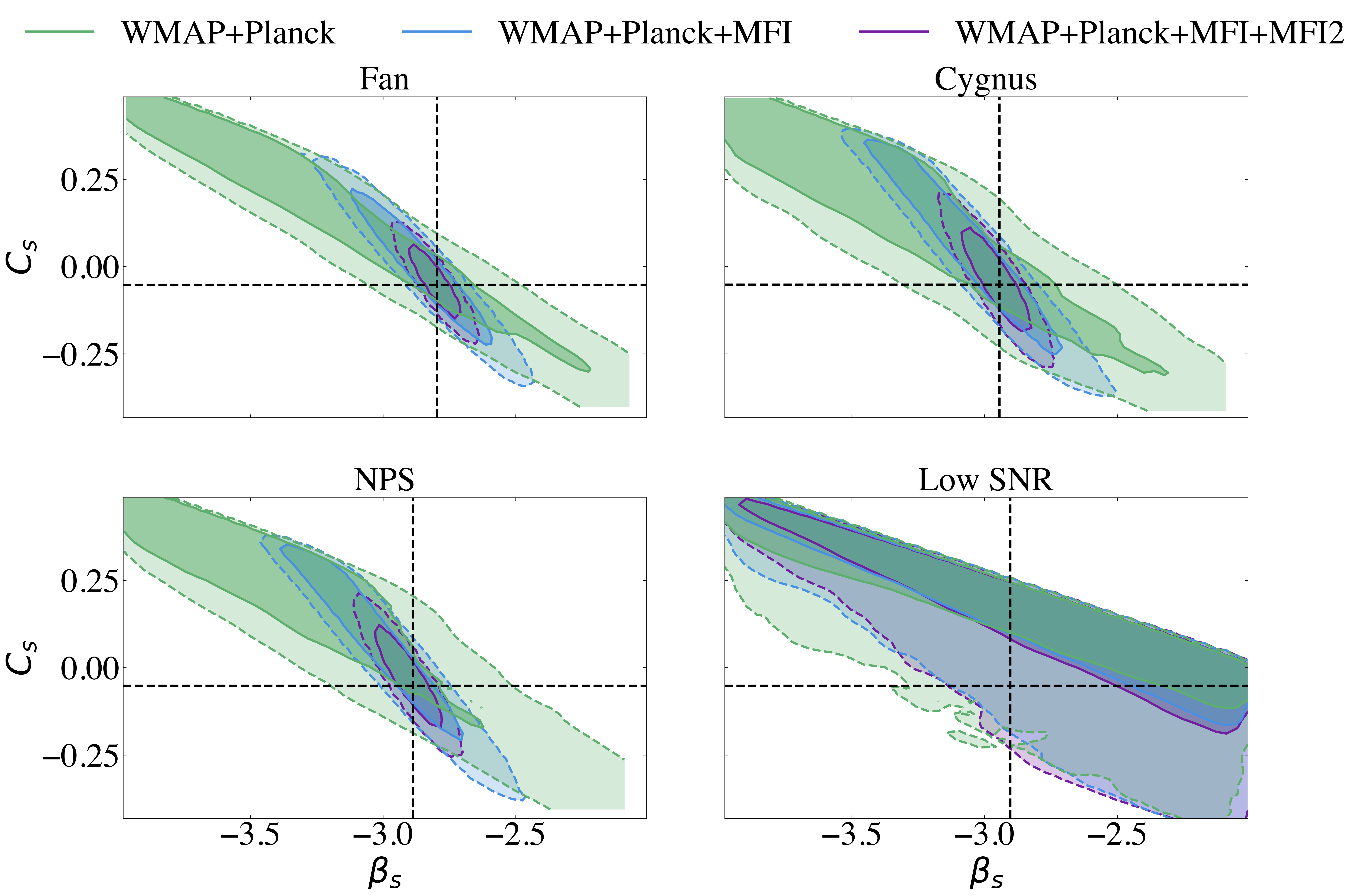}
    \caption{2D marginalised PDFs of the synchrotron spectral indices ($\beta_s$, $C_s$), for the four case-study pixels in the wide survey configuration, shown for WMAP+\textit{Planck}, WMAP+\textit{Planck}+MFI and WMAP+\textit{Planck}+MFI+MFI2 datasets. Contours indicate the 68\% and 95\% posterior density regions, with the true parameter values being marked by the black dashed lines.}
    \label{fig:2djoint_curv_ws}
\end{figure}

The noise dominated pixel reveals a different picture. The synchrotron spectral index $\beta_s$ is effectively unconstrained by all datasets, as seen in Fig. \ref{fig:posteriors_no_curv_ws}, with the posteriors being completely prior-dominated. Consequently, the improvement factor for $\beta_s$ is modest, about 1.3 relative to WMAP+\textit{Planck} (Table \ref{tab:improv_f}), reflecting only a slight reduction in uncertainties and bias with the addition of QUIJOTE-MFI and MFI2. Because the posterior is prior-dominated, however, this factor does not meaningfully indicate improved constraints on $\beta_s$. By contrast, the amplitudes $A_{Q,s}$ and $A_{U,s}$ benefit substantially from QUIJOTE data, with total errors reduced by factors of $\sim$3–4.

Bayesian or parametric measurements in these regions are largely prior-dominated \citep{2020A&A...641A...4P, MFIcompsep_pol}, while patch-averaged T–T plots \citep{2014ApJ...790..104F} or template fitting \citep{2022MNRAS.513.5900H, 2019MNRAS.485.2844D} reduce uncertainties by averaging over many pixels under the assumption of a constant $\beta_s$, but individual pixels remain poorly constrained and results are sensitive to method, template choice, and residual foregrounds. As will be discussed in Section \ref{sec:results_pixels_cf}, the QUIJOTE-MFI2 cosmological fields overcomes these limitations.

\subsubsection{Curved synchrotron} \label{sec:results_pixels_ws_curv}

We now extend our forecasts to a curved parameterisation of the synchrotron spectrum (Eq. \ref{eq:model_sync_curv}) within the wide survey. The marginalised PDFs are shown in Fig. \ref{fig:posteriors_curv_ws} and the corresponding improvement factors can be found in Table \ref{tab:improv_f}.

First considering the high SNR pixels, the inclusion of MFI2 leads to a substantial reduction of the posterior widths of all four synchrotron parameters. The synchrotron amplitude in Q, $A_{Q,s}$ (amplitude in U, $A_{U,s}$) decreases from $\sigma \sim$0.29-0.35 $\mathrm{mK_{RJ}}$ ($\sim$ 0.06-0.12 $\mathrm{mK_{RJ}}$) with WMAP+\textit{Planck}, to $\sigma \sim$0.04 $\mathrm{mK_{RJ}}$ ($\sim$0.02-0.03 $\mathrm{mK_{RJ}}$) with MFI, and further to $\sigma \sim$ 0.01 $\mathrm{mK_{RJ}}$ (0.01 $\mathrm{mK_{RJ}}$) with MFI2. Constraints on the spectral parameters are also tightened: $\beta_s$, which is only loosely constrained by WMAP+\textit{Planck} with uncertainties exceeding 0.5, is narrowed to about 0.16-0.2 with MFI, and reaches the 0.07-0.09 level with MFI2. For $C_s$, uncertainties of 0.28 with WMAP+\textit{Planck} are reduced to 0.13-0.17 with MFI and to below $0.1$ with MFI2.

The introduction of curvature allows for more flexibility in describing the synchrotron emission, but it also makes the estimation of the spectral parameters more prone to biases. As seen in the biased marginal PDFs in Fig. \ref{fig:posteriors_curv_ws} and the normalised bias values reported in Table \ref{tab:bias_curv}, WMAP+\textit{Planck} data alone exhibit systematic biases across all parameters and high SNR pixels: for example, $|\mathrm{bias}/\sigma|$ reaches 0.67-0.77 for $\beta_s$ and 0.78-0.96 for $C_s$. The addition of MFI2 reduces these biases to levels that are negligible compared to the posterior uncertainties, with values reduced to $|\mathrm{bias}(\beta_s)/\sigma(\beta_s)| \lesssim$ 0.16 for $\beta_s$ and $|\mathrm{bias}(C_s)/\sigma(C_s)| \lesssim$ 0.22 for $C_s$. The bias distributions of $\beta_s$ and $C_s$ over a thousand realisations, shown in Fig. \ref{fig:bias_hist_BsCs_curv}, are narrowly centred around zero with MFI2, in contrast to the broader, systematically offset distributions from WMAP+\textit{Planck}. Although the biases at the pixel level are below 1$\sigma$, because they are systematic they do not average out over larger sky areas (for example when estimating a mean $\beta_s$), and therefore can become statistically significant at the map level. In other words, unbiased recovery of the spectral parameters is only statistically achieved with MFI2.

The reduction in the uncertainties of the synchrotron parameters, together with the effective removal of biases achieved with QUIJOTE-MFI and MFI2, is reflected in the improvement factors listed in Table \ref{tab:improv_f}: approximately $8-10$ for $\beta_s$, and $33-43$ and $6-14$ for the $Q$ and $U$ amplitudes respectively, in high SNR regions (Table \ref{tab:improv_f}). For parameters such as $C_s$, where the WMAP+\textit{Planck} posterior is fully dominated by the prior, the nominal improvement factor of $3-5$ fails to convey the full significance of adding QUIJOTE-MFI and MFI2, which provide the first substantial constraints on this parameter beyond the prior. These improvement factors are consistently higher than those obtained for the pure power-law case in Section \ref{sec:results_pixels_ws_pl}, reflecting the overall reduction in bias achieved by the inclusion of MFI2 across all parameters.

\begin{figure}
    \centering
    \includegraphics[width=\columnwidth]{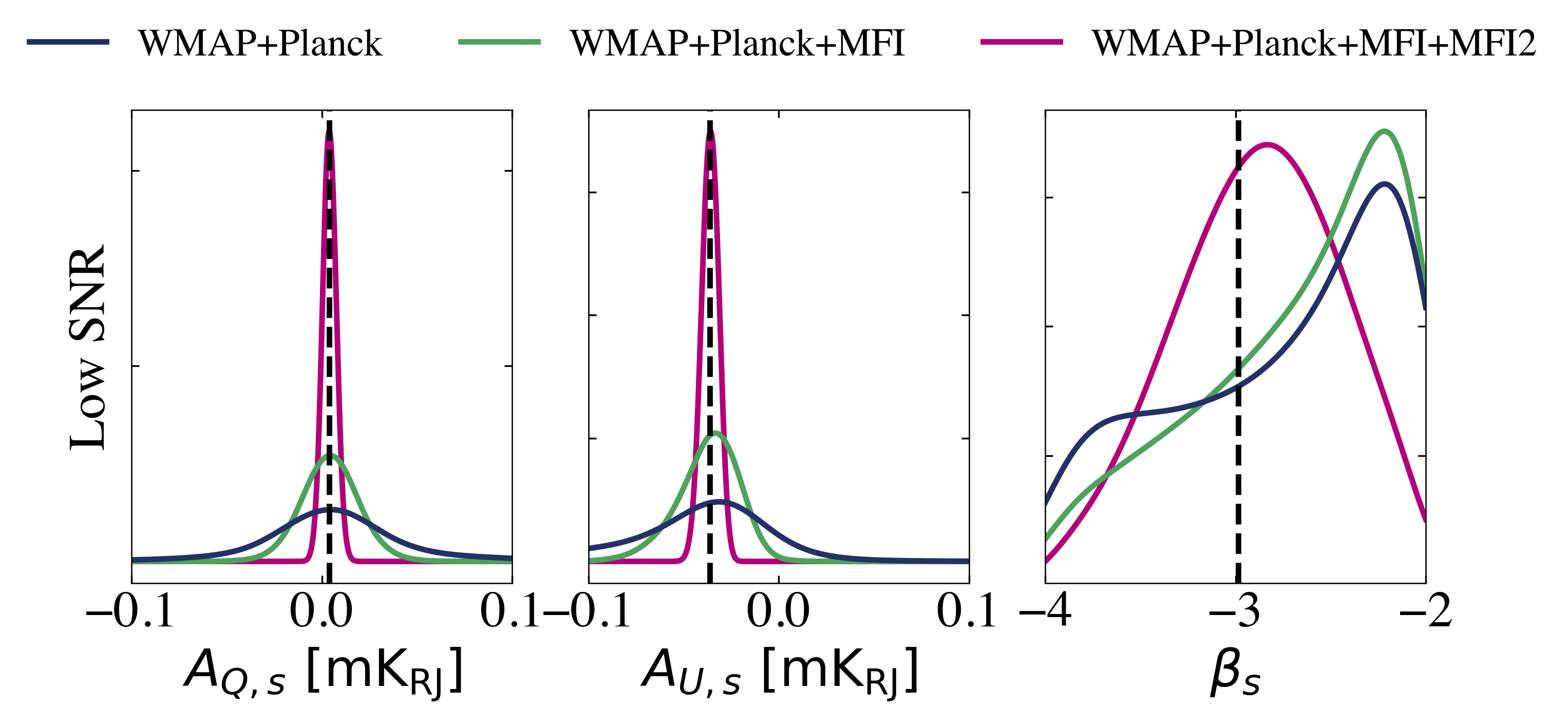}
    \caption{1D marginalised PDFs for the synchrotron parameters in the low SNR pixel, for the cosmological fields forecast, assuming the synchrotron power-law model, shown for WMAP+\textit{Planck} (blue), WMAP+\textit{Planck}+MFI (green) and WMAP+\textit{Planck}+MFI+MFI2 (pink) datasets. The black dashed lines are the true parameter values known from \texttt{PySM}.}
    \label{fig:posteriors_no_curv_lowsnr_cf}
\end{figure}

\begin{figure}
    \centering
    \includegraphics[width=\columnwidth]{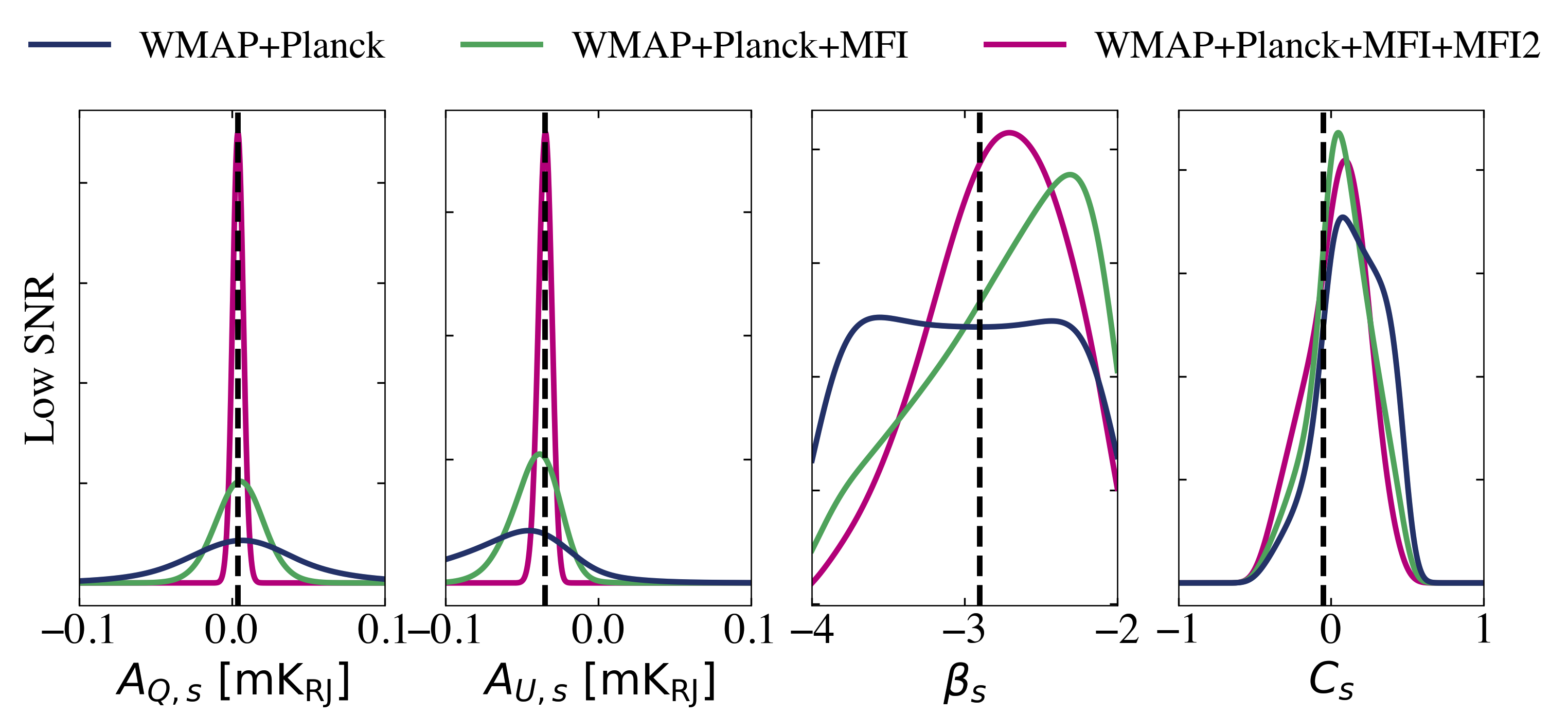}
    \caption{1D marginalised PDFs for the synchrotron parameters in the low SNR pixel, for the cosmological fields forecast, assuming curved synchrotron spectrum. The black dashed lines are the true parameter values known from \texttt{PySM}.}
    \label{fig:posteriors_curv_cf}
\end{figure}

Moreover, as the synchrotron spectrum departs from a strict power-law behaviour, a degeneracy emerges between $\beta_s$ and $C_s$ that is challenging to resolve, as illustrated in Fig. \ref{fig:2djoint_curv_ws}. In particular, the parameters exhibit a clear anticorrelation, with higher values of $\beta_s$ corresponding to lower values of $C_s$. Even with the additional low-frequency information provided by QUIJOTE-MFI2, the correlation between these parameters cannot be fully broken, despite substantial improvements in the high SNR pixels.

Given the degeneracy between $\beta_s$ and $C_s$ and the resulting broadened posteriors, a key question is whether the data allow for a statistically significant detection of synchrotron curvature. For example, in the Fan pixel, the recovered curvature with QUIJOTE-MFI and MFI2 is $ C_s = -0.044_{-0.073}^{+0.072}$,
consistent with the input value of $-0.052$ but also consistent with zero. This illustrates that, even with WMAP+\textit{Planck}+MFI+MFI2, a curvature of order $|C_s|\simeq0.05$ cannot be robustly distinguished from a simple power-law. In practice, only for $|C_s|\gtrsim0.14$ would the data permit a $2\sigma$ detection of curvature in an individual high SNR pixel such as the Fan.

In the low SNR pixel, the recovery of synchrotron parameters remains inherently challenging due to large uncertainties and biases. The posteriors for $\beta_s$ and $C_s$ remain fully dominated by the prior (see the bottom panel of Fig.~\ref{fig:posteriors_curv_ws}), and the improvement factors listed in Table~\ref{tab:improv_f} are modest. These results indicate that, within the QUIJOTE-MFI and MFI2 frequency range, the wide-survey sensitivity is insufficient to meaningfully constrain the synchrotron spectrum in low SNR regions on a pixel-by-pixel basis, and residual biases in both $\beta_s$ and $C_s$ indicate that these parameters cannot be reliably recovered. Since the pivot frequency of the synchrotron model is at 10.35 GHz, curvature effects are primarily constrained by data at frequencies away from this scale, and constraining $C_s$ in low SNR regions requires broader frequency coverage than provided by the current QUIJOTE bands.

We also tested whether the synchrotron models can be meaningfully distinguished in practice. We evaluated a random realisation from the ensemble using $\chi^2$ statistics. Fits with both the correct and incorrect spectral model yielded $\chi^2$ values between 0.92 and 0.96. Consequently, in the context of real data, where the true underlying spectrum is unknown, for a curvature level $C_s=-0.052$ both power-law and curved models would provide statistically acceptable fits.

\subsection{Cosmological fields results} \label{sec:results_pixels_cf}

We next present forecasts for the cosmological fields, which are located in intrinsically low-brightness regions of the sky. The analysis focuses on a representative low SNR pixel, where CMB polarisation recovery is particularly sensitive to foreground separation. As in the wide-survey forecast, we first adopt a simple power-law model for the synchrotron spectrum, then extend the analysis to include spectral curvature.

\subsubsection{Power-law synchrotron} \label{sec:results_pixels_cf_pl}

Figure \ref{fig:posteriors_no_curv_lowsnr_cf} shows the marginalised PDFs for the cosmological field, power-law synchrotron case. As discussed in Section \ref{sec:results_pixels_ws_pl}, constraining the spectral index $\beta_s$ in regions of intrinsically low SNR has long been a challenge. In the low SNR pixel considered here, WMAP+\textit{Planck} alone provide no meaningful constraints on $\beta_s$, yielding a posterior fully dominated by the prior ($\sigma(\beta_s) \sim 0.74$, bias 0.32). The inclusion of MFI data does not change this substantially, with the posterior remaining prior-driven ($\sigma \sim 0.67$, bias 0.38). Only when MFI2 data are added does $\beta_s$ become genuinely constrained by the data, with $\sigma(\beta_s) \sim 0.59$ and bias reduced to 0.29. This marks a clear transition from prior-limited to data-driven inference. While an improvement factor of 1.8 can be quoted relative to WMAP+\textit{Planck}, it should be interpreted with caution, since the reference uncertainty is not data-defined. The polarisation amplitudes $A_{s,Q}$ and $A_{s,U}$ also show substantial gains, with improvement factors of 10 (Table \ref{tab:improv_f}), comparable to those obtained for the wide survey.

\subsubsection{Curved synchrotron} \label{sec:results_pixels_cf_curv}

Finally, we extend the forecasts for the cosmological fields to a curved synchrotron spectrum, representing the most challenging scenario due to the combination of low intrinsic brightness and increased spectral complexity. Figure \ref{fig:posteriors_curv_cf} presents the marginalised PDFs for the low SNR pixel.

In this case, WMAP+\textit{Planck} alone, and even WMAP+\textit{Planck} combined with MFI, produce biased estimates for all synchrotron parameters. For the synchrotron amplitudes, MFI2 is able to make unbiased measurements, while also reducing the statistical uncertainties to $0.003$–$0.004\,\mathrm{mK}_{\mathrm{RJ}}$. This translates into improvement factors of 11 relative to WMAP+\textit{Planck}.

Even with the introduction of spectral curvature in the synchrotron modelling, MFI2 remains able to constrain the synchrotron spectral index $\beta_s$, whereas WMAP+\textit{Planck} alone provide little information and WMAP+\textit{Planck}+MFI remain prior dominated. With MFI2, the posterior becomes data-driven, with $\sigma(\beta_s) \sim$ 0.42, corresponding to an improvement factor of 1.6 over WMAP+\textit{Planck}. The curvature parameter $C_s$ continues to be strongly shaped by the prior across datasets. This indicates that, while the improved sensitivity and band sampling of MFI2 are sufficient to render the synchrotron spectral index $\beta_s$ data-driven, they do not yet provide enough low-frequency leverage to fully break the intrinsic degeneracy between $\beta_s$ and $C_s$ on a pixel-by-pixel basis. Nevertheless, the addition of MFI2 reduces the nominal bias, and uncertainty, from $\sigma(C_s) \sim 0.28$ with WMAP+\textit{Planck} to $\sigma(C_s) \sim 0.22$, corresponding to a formal improvement factor of 1.2.

Finally, even though the low-frequency QUIJOTE data points alleviate somewhat the degeneracy between $\beta_s$ and $C_s$, it still persists. Achieving a full decorrelation between these parameters would require either additional low-frequency channels to further constrain the synchrotron spectrum, or complementary data at intermediate frequencies to better separate curvature effects from the spectral index. Nonetheless, the current forecasts demonstrate that MFI2, in combination with WMAP and \textit{Planck}, provides a substantial step forward in constraining both the amplitude and spectral parameters of synchrotron emission in the low SNR regions.

\begin{figure}
    \centering
    \includegraphics[width=\columnwidth]{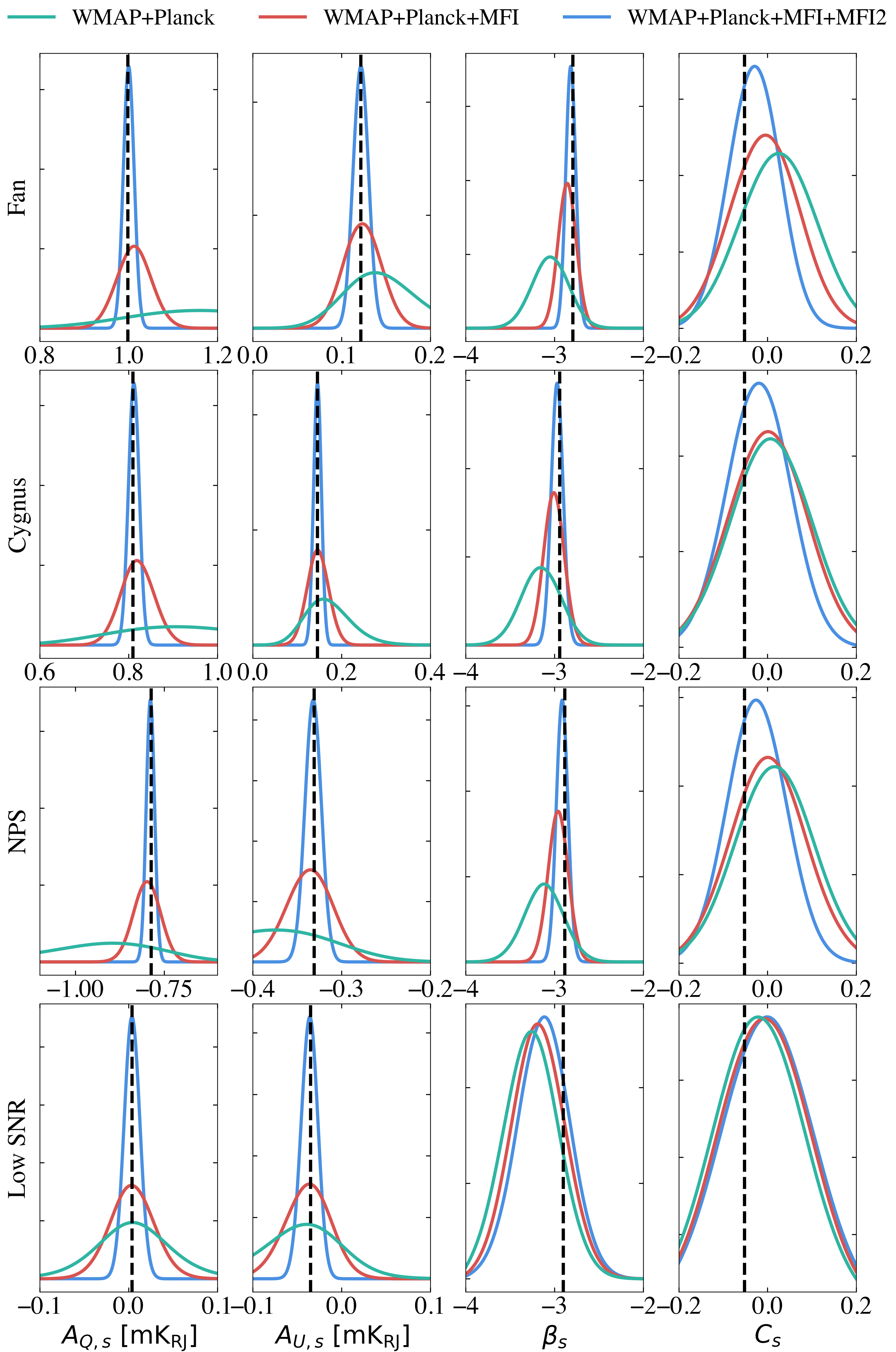}
    \caption{1D marginalised PDFs for the synchrotron parameters for the wide survey assuming a curved synchrotron spectrum, obtained using Gaussian priors. Colours indicate the datasets: WMAP+\textit{Planck} (green), WMAP+\textit{Planck}+MFI (orange), and WMAP+\textit{Planck}+MFI2 (blue). Black dashed lines show the true \texttt{PySM} values.}
    \label{fig:posteriors_curv_gaussian}
\end{figure}

\section{Results: map level forecasts} \label{sec:results_map}

Here we extend the analysis of the case study pixels to the full survey and the deep cosmological fields. The MFI wide survey covers approximately 29,000\,$\mathrm{deg}^2$, corresponding to the full available sky after applying the QUIJOTE MFI default mask, which excludes the geostationary satellites, North Celestial Pole, and low declination regions. We also forecast for the three QUIJOTE cosmological fields, each targeting low SNR regions with an area of approximately 1,000 $\mathrm{deg}^2$ per field. We compute all results on pixels of size $N_{\rm side}=64$ with a Gaussian beam of $1^\circ$ FWHM, which provides a consistent resolution for parameter estimation and prevents artificially small uncertainties for parameters that are nearly uniform across the sky, such as $C_s$. Results are presented separately for the wide survey configuration and the cosmological field patches, and for both the power-law and curved synchrotron models.

The synchrotron spectral index is prior-dominated in areas of the sky with weak synchrotron emission, as seen in the pixel-level study (Section \ref{sec:results_pixels}), where it remained mostly unconstrained under a Jeffreys prior. In full sky forecasts this poses a challenge: in such regions, non informative priors such as Jeffreys provide little meaningful constraints, leading to unreliable synchrotron parameter recovery. Moreover, in studies focusing on low SNR regions, the accurate recovery of CMB amplitudes is the main objective, while the synchrotron emission is often below detection levels. For this reason, informative Gaussian priors on the synchrotron spectral index are commonly adopted, which assume that the measured value in high SNR regions extends to the low SNR ones. To emulate this methodology, we perform the full sky forecast with Gaussian priors rather than Jeffreys. To enable a direct comparison with previous component separation studies based on real QUIJOTE-MFI data \citep{MFIcompsep_pol}, we adopt the same Gaussian priors on all spectral indices: for the synchrotron spectral index $\beta_s$ we use $\mathcal{N}(-3.1, 0.3)$, for the synchrotron curvature $C_s$ we use $\mathcal{N}(0, 0.1)$, for the dust spectral index $\beta_d$ we use $\mathcal{N}(1.55, 0.1)$, and for the dust temperature $T_d$ we use $\mathcal{N}(21, 3)$ K. 

To illustrate the impact of adopting Gaussian priors, Figure \ref{fig:posteriors_curv_gaussian} shows the marginalised PDFs for the four case study pixels in the wide survey configuration under the curved synchrotron model. This case is representative, and we restrict the comparison to it for clarity. Comparing the posteriors obtained under Gaussian and Jeffreys priors (Fig. \ref{fig:posteriors_curv_ws}) reveals several systematic differences across the four case study pixels. These differences are most apparent in two situations: (i) in the low SNR pixel, where there is little signal to constrain the synchrotron spectral indices, and the apparent constraints for the three datasets are entirely prior driven. Under Jeffreys priors, the posteriors remain unconstrained ($\sigma(\beta_s)\sim$0.55 for MFI+MFI2 alone, $\sim$0.68 for WMAP+\textit{Planck}), whereas Gaussian priors yield well-defined Gaussian posteriors with reduced uncertainties matching the prior width ($\sigma(\beta_s)\sim$0.3). (ii) In the high SNR pixels with WMAP+\textit{Planck} alone, the posteriors remain sensitive to the choice of prior despite the higher SNR. Adding MFI reduces this sensitivity but does not remove it entirely. However, with the full combination WMAP+\textit{Planck}+MFI+MFI2, the data become sufficiently constraining that the posteriors are effectively independent of the Gaussian prior.

\subsection{Wide survey results} \label{sec:results_map_ws} 

We begin with the wide survey observing mode, modelling the synchrotron emission as a power-law spectrum.

\begin{figure*}
    \centering
    \includegraphics[width=\textwidth]{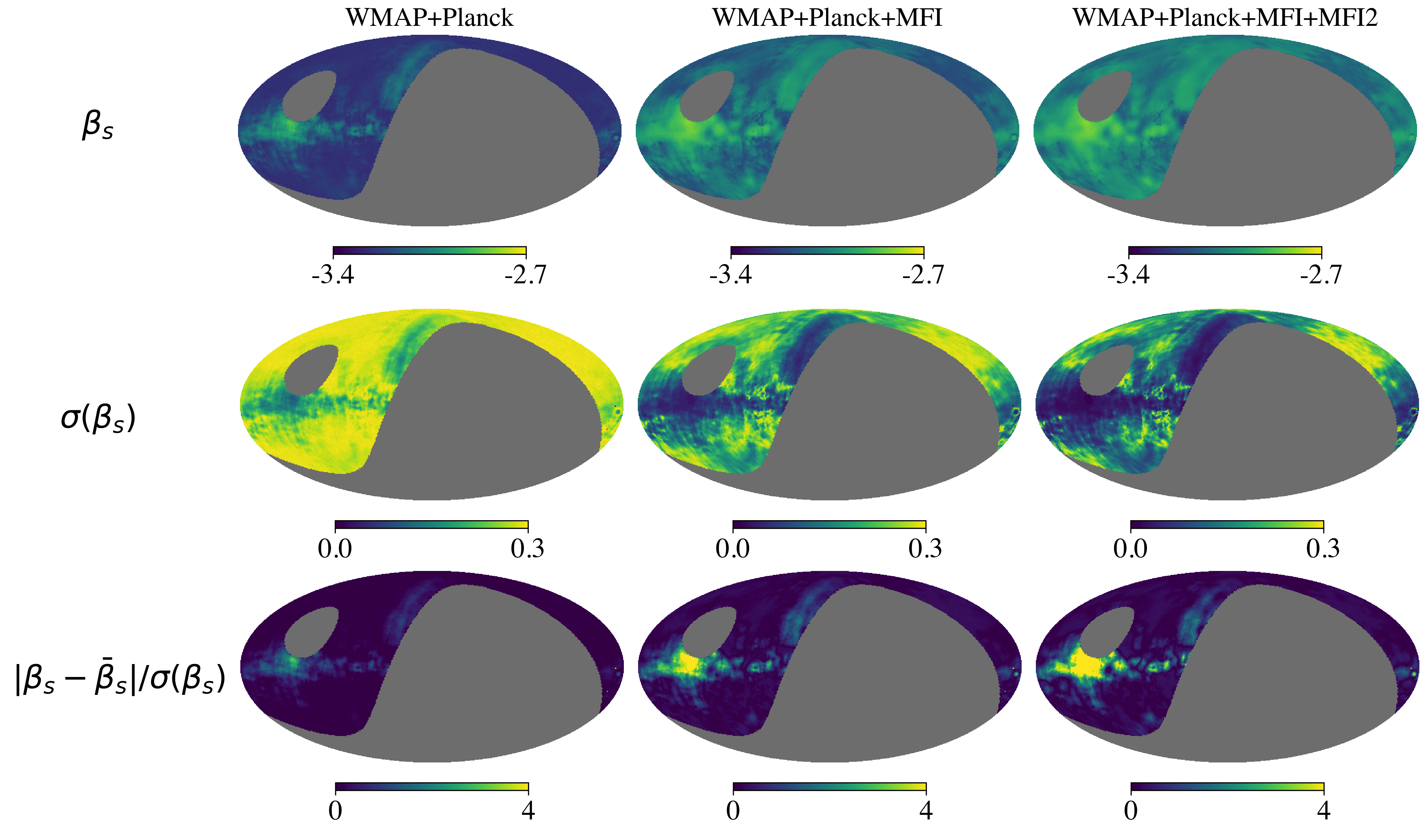}
    \caption{Maps of $\beta_s$ in the wide survey assuming a synchrotron power-law model. Columns show the three datasets: WMAP+\textit{Planck} (left), WMAP+\textit{Planck}+MFI (centre), and WMAP+\textit{Planck}+MFI+MFI2 (right). Rows correspond to the recovered median $\beta_s$ (top), its associated uncertainty (middle) and the normalised deviation relative to the median of unmasked pixels $\bar{\beta}_s$, defined as $|\beta_s - \bar{\beta}_s| / \sigma(\beta_s)$ (bottom). Grey areas indicate the masked pixels. These maps show the recovered quantities and highlight relative spatial variations. The bottom row quantifies how strongly each pixel deviates from the map median, rather than from the true input $\beta_s$.}
    \label{fig:map_Bs_no_curv_ws}
\end{figure*}

\subsubsection{Power-law synchrotron}  \label{sec:results_map_ws_pl}

The reconstructed median maps of $\beta_s$, their associated uncertainties $\sigma(\beta_s)$ and the ratio $|\beta_s - \bar{\beta}_s| / \sigma(\beta_s)$ ($\bar{\beta}_s$ being the map median), for the wide survey power-law case are shown in Fig. \ref{fig:map_Bs_no_curv_ws} for the three datasets. These maps show recovered quantities from the component separation analysis. In particular, the normalised deviation shown in the bottom plot quantifies the significance of spatial variations with respect to a global reference value, rather than agreement with the input model.

Most of the sky remains prior-dominated, with the recovered uncertainty on the synchrotron spectral index remaining close to the prior width of 0.3 over large fractions of the sky, as expected given the weak synchrotron signal in many regions. This is particularly the case for WMAP+\textit{Planck}, for which $\beta_s$ remains effectively unconstrained over much of the sky. In these low SNR regions, the recovered $\beta_s$ can vary between datasets, especially where the true values deviate from the prior, reflecting the combined effect of dataset-specific fluctuations and limited constraining power. The inclusion of QUIJOTE-MFI and MFI2 leads only to modest improvements in these regions.

In contrast, in regions of strong synchrotron emission such as the Galactic plane ($|b| < 10^\circ$), the recovery of the spectral index is data-driven for all datasets, with similar median values close to the prior mean of $-3.1$. While the inclusion of QUIJOTE-MFI already reduces the median uncertainty on $\beta_s$ from 0.23 for WMAP+\textit{Planck} alone to 0.10, the addition of MFI2 provides a further substantial improvement, lowering the uncertainty to 0.06. The significance of spatial variations, quantified as $|\beta_s - \bar{\beta}_s| / \sigma(\beta_s)$, also increases with the inclusion of low-frequency data. This increase reflects the enhanced ability of MFI and MFI2 to detect spatial variations: the typical significance of deviations relative to the map median rises from 0.07 (WMAP+\textit{Planck}) to 0.26 (+MFI2), and the fraction of pixels exceeding 1$\sigma$ grows from 4.4\% to 16.4\%, showing that low-frequency data improve the sensitivity to local departures from the median spectral index, particularly in regions of strong emission, while most of the sky remains consistent with a single global value. 

Comparison of the forecasted $\sigma(\beta_s)$ maps with those derived from real WMAP+\textit{Planck}+MFI component-separated data \citep{MFIcompsep_pol} shows broadly consistent uncertainty levels, with most of the sky being dominated by the prior width of 0.3. The analysis of \citet{MFIcompsep_pol} was performed at a resolution of $2^{\circ}$, which naturally yields smaller per-pixel uncertainties than in this forecast. The forecasted $\beta_s$ maps underestimate the spatial scatter of the synchrotron spectral index, reflecting the tendency of \texttt{PySM} models to underestimate the true spatial dispersion of $\beta_s$. This discrepancy is largest in the Galactic plane, where the simple power-law model cannot capture the full complexity, while low SNR regions, being prior-dominated, are well reproduced.

\subsubsection{Curved synchrotron} \label{sec:results_map_ws_curv}

\begin{figure*}
    \centering
    \includegraphics[width=\textwidth]{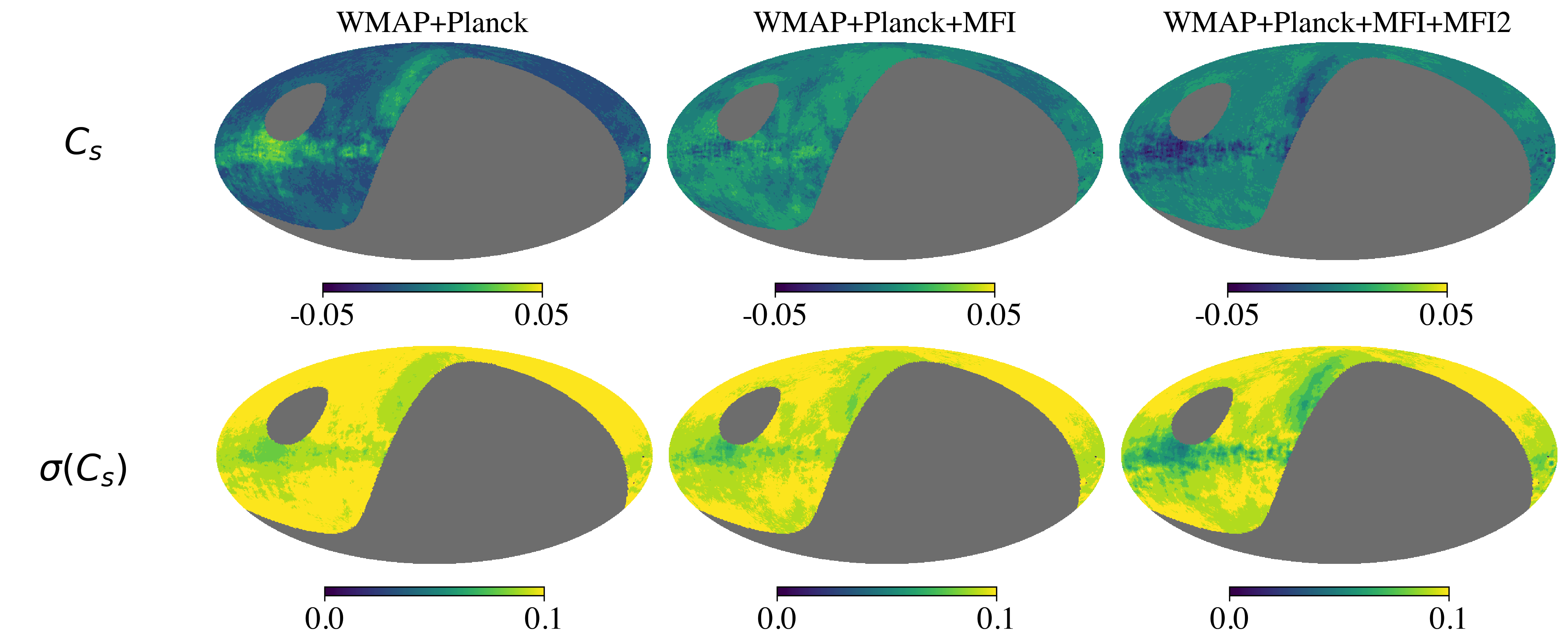}
    \caption{Maps of $C_s$ in the wide survey assuming a synchrotron power-law model. Columns show the three datasets: WMAP+\textit{Planck} (left), WMAP+\textit{Planck}+MFI (centre), and WMAP+\textit{Planck}+MFI+MFI2 (right). Rows correspond to the recovered median $C_s$ (top) and its associated uncertainty (bottom). Grey areas indicate the masked pixels.}
    \label{fig:map_Cs_curv_ws}
\end{figure*}

Continuing with the wide survey analysis, we now turn to the case of a curved synchrotron spectrum. Figure \ref{fig:map_Cs_curv_ws} presents the full sky constraints for the curvature spectral index $C_s$. 
The corresponding $\beta_s$ maps are not shown, as their interpretation closely mirrors the previous power-law case.

The addition of QUIJOTE-MFI and MFI2 improves constraints on synchrotron curvature primarily in the brightest regions of the Galactic plane, while high-latitude and faint-sky regions remain dominated by the prior. Forecasted uncertainties decrease from $\sigma(C_s) = 0.091$ with WMAP+\textit{Planck} to 0.087 with MFI and further to 0.078 with MFI2, raising the mean $|C_s|/\sigma(C_s)$ in the Galactic plane from 0.12 to 0.16. In most low-emission regions, $C_s$ is prior-limited ($\sigma \sim 0.1$) and consistent with zero.

At high Galactic latitudes, $|C_s|/\sigma(C_s)$ remains below $\sim$0.06, and even in the brightest synchrotron regions such as the Fan and the NPS it reaches at most unity, leaving the input curvature of the s3 model ($C_s = 0.052$) undetectable at the per-pixel level. A 2$\sigma$ detection of curvature on a single pixel would require $|C_s| \sim 0.18$. Because the input curvature is spatially constant in these forecasts, averaging over extended, high-emission regions can recover $C_s$.

These forecasts are consistent with \citet{MFIcompsep_pol}, which found that curvature is only detected in limited parts of the Galactic plane at $\sim 3\sigma$ (with amplitudes significantly larger than those adopted in the s3 model), while at high latitudes the synchrotron spectrum is consistent with a pure power law and $C_s$ remains unconstrained. In that work, a measurement of curvature was achieved by averaging over large regions at a resolution of $2^{\circ}$, under the assumption of a constant $C_s$, with larger pixels naturally providing greater sensitivity. This analysis represents a natural extension of the work presented here, but it lies beyond the scope of this paper.

Finally, as in \citet{MFIcompsep_pol}, we could also evaluate the reduced $\chi^2$, or another goodness-of-fit statistic, to perform model selection and compare fits with and without the curvature parameter. Given the intrinsic curvature of the s3 model and the typical uncertainties discussed above, we find that the reduced $\chi^2$ is nearly identical in both cases (0.95 and 0.96, respectively). This implies that, for this simulation, a model without curvature is sufficient.

\begin{figure*}
    \centering
    \includegraphics[width=\textwidth]{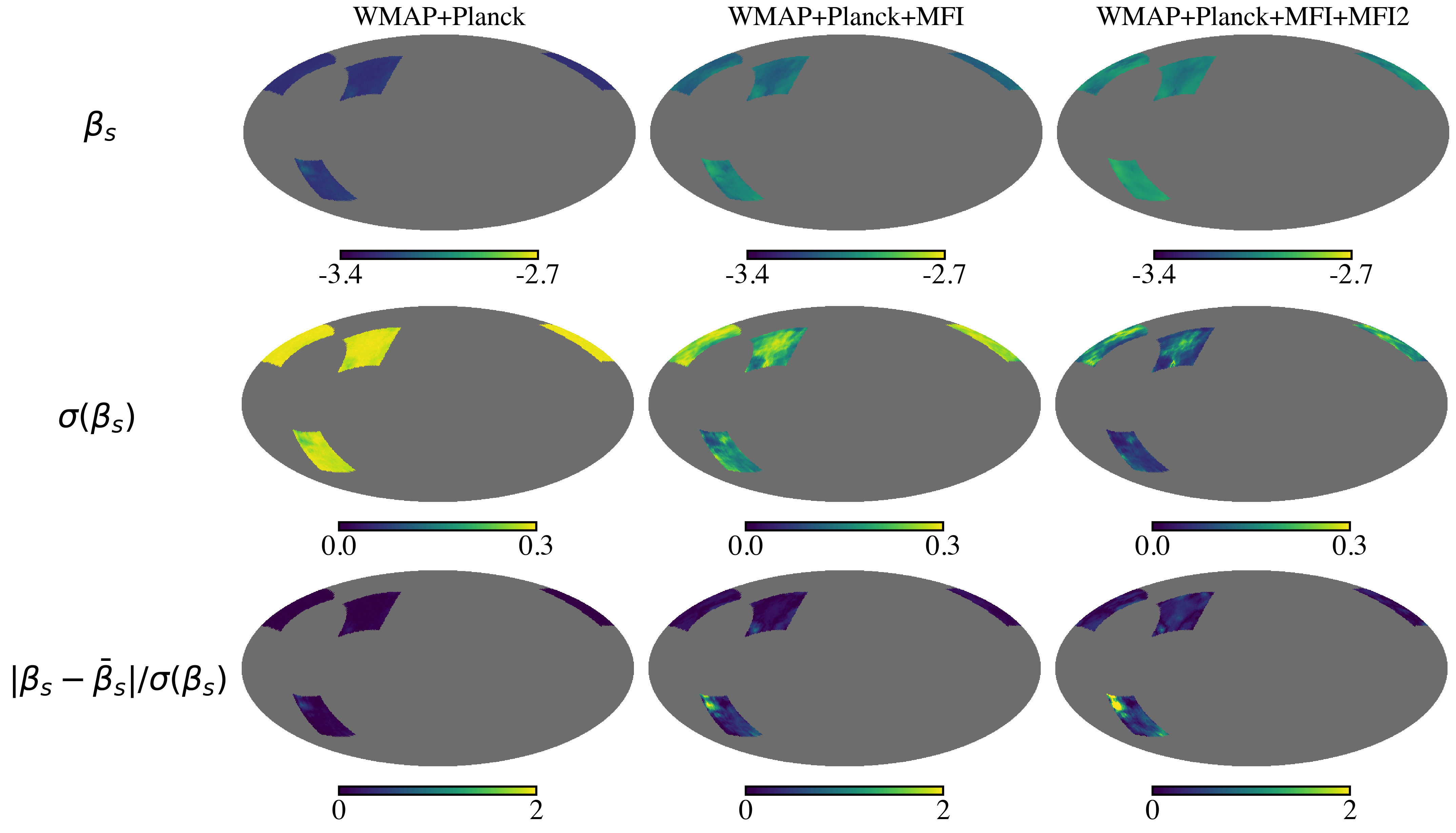}
    \caption{Maps of $\beta_s$ in the cosmological fields assuming a synchrotron power-law model. Columns show the three datasets: WMAP+\textit{Planck} (left), WMAP+\textit{Planck}+MFI (centre), and WMAP+\textit{Planck}+MFI+MFI2 (right). Rows correspond to the recovered median $\beta_s$ (top), its associated uncertainty (middle) and the SNR relative to the median of unmasked pixels, $|\beta_s - \bar{\beta}_s| / \sigma(\beta_s)$ (bottom). Grey areas indicate the masked pixels.}
    \label{fig:map_Bs_no_curv_cf}
\end{figure*}

\subsection{Cosmological fields results}  \label{sec:results_map_cf}

Having discussed the wide survey, we now present the forecast for the QUIJOTE cosmological fields.

\subsubsection{Power-law synchrotron}  \label{sec:results_map_cf_pl}

Forecasts of the synchrotron spectral index $\beta_s$ in the QUIJOTE deep cosmological fields are presented for a power-law synchrotron model in Figure \ref{fig:map_Bs_no_curv_cf}.

In the deep cosmological fields, WMAP+\textit{Planck} alone provide limited constraints on the synchrotron spectral index, with a mean uncertainty of 0.29, with reliable constraints achieved in just 1.5\% of the pixels. Adding MFI improves the recovery moderately, lowering the mean uncertainty to 0.21, but a significant fraction of the field remains poorly constrained. Including MFI2 results in a clear improvement in uncertainty, bringing the mean down to 0.12.

\begin{figure}
    \centering
    \includegraphics[width=\columnwidth]{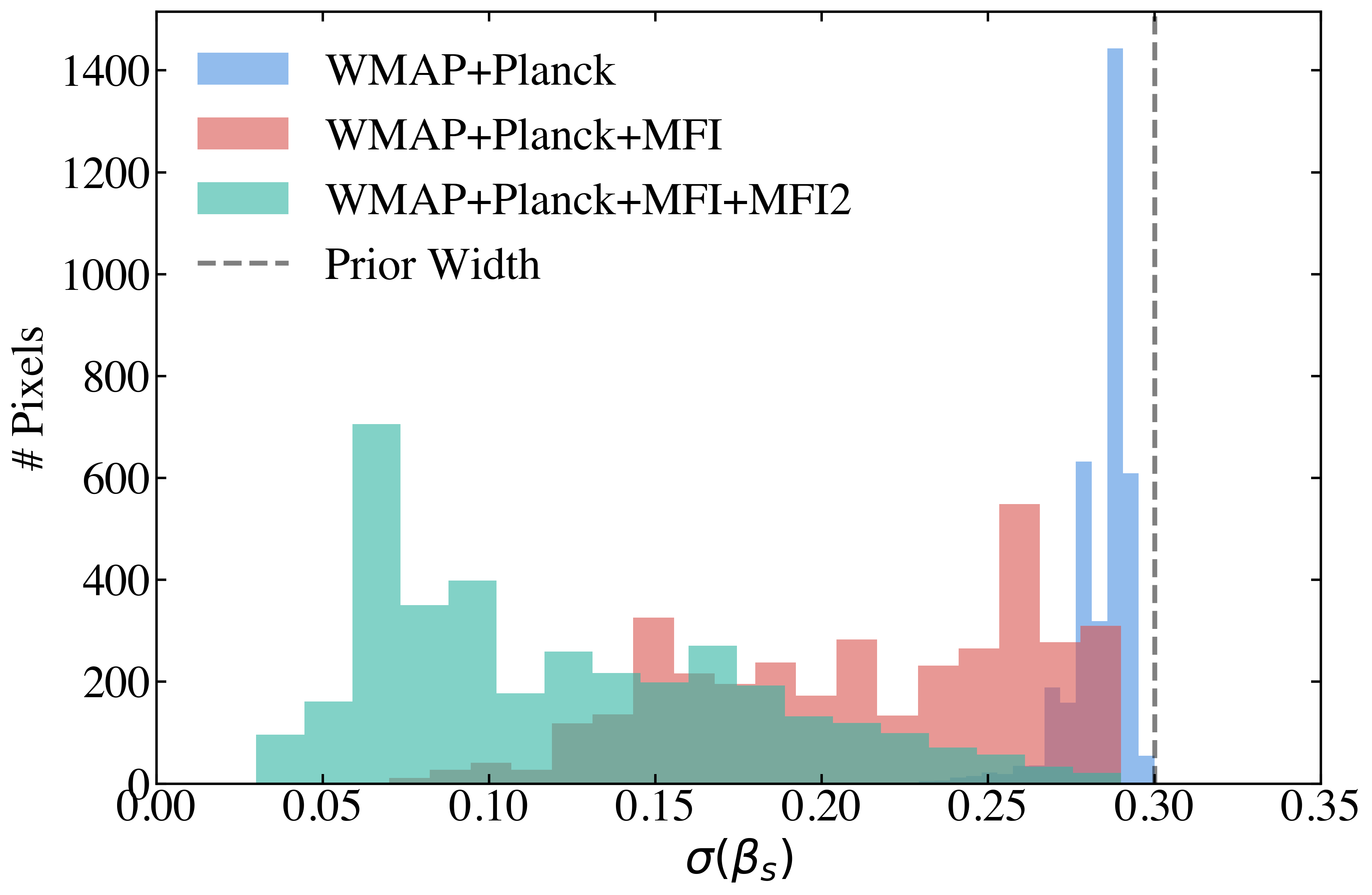}
    \caption{Distributions of the forecasted uncertainties on the synchrotron spectral index, $\sigma(\beta_s)$, across the three QUIJOTE cosmological fields for the WMAP+\textit{Planck} (blue), WMAP+\textit{Planck}+MFI (orange), and WMAP+\textit{Planck}+MFI+MFI2 (green) data combinations. The grey dashed line indicates the upper limit imposed by the width of the Gaussian prior on $\beta_s$.}
    \label{fig:histogram_sigma_beta_s_cf}
\end{figure}

The histogram of $\sigma(\beta_s)$ (Figure \ref{fig:histogram_sigma_beta_s_cf}) further illustrates this impact: MFI2 shifts the distribution toward lower uncertainties and greatly increases the number of pixels with robust spectral index estimates. In contrast to the wide survey, where only $\sim$ 350 of 3560 pixels are meaningfully constrained in these low SNR regions, the deep-field mode of MFI2 provides reliable $\beta_s$ constraints for $\sim$ 3506 pixels. The bottom panel of Figure \ref{fig:map_Bs_no_curv_cf} shows maps of $|\beta_s - \bar{\beta_s}| / \sigma(\beta_s)$, quantifying the per-pixel significance of spatial deviations from the median. Including MFI and MFI2 progressively reveals more significant variations, with the fraction of pixels exceeding $1\sigma$ increasing from 0\% for WMAP+\textit{Planck} to 5.7\% for WMAP+\textit{Planck}+MFI+MFI2.

Moreover, in order to assess the level of foreground contamination at CMB frequencies, we propagate the forecasted uncertainties on the synchrotron amplitudes to 100 GHz, selected as a representative cosmological band near the foreground minimum where CMB measurements are typically performed. For each pixel, we evaluate the synchrotron emission at 100 GHz for every posterior sample of the MCMC chains, with the residual uncertainty defined as the standard deviation of the resulting posterior distribution. This approach fully accounts for correlations between the synchrotron amplitude and spectral index, as well as the non linear frequency scaling of the model. For WMAP+\textit{Planck} alone, the median synchrotron residual at 100 GHz is 0.21 \textmu K$_{\mathrm{CMB}}$. The inclusion of MFI and MFI2 reduces this residual to 0.033 \textmu K$_{\mathrm{CMB}}$, representing roughly a sixfold reduction relative to WMAP+\textit{Planck} alone.

\subsubsection{Curved synchrotron} \label{sec:results_map_cf_curv}

\begin{figure*}
    \centering
    \includegraphics[width=\textwidth]{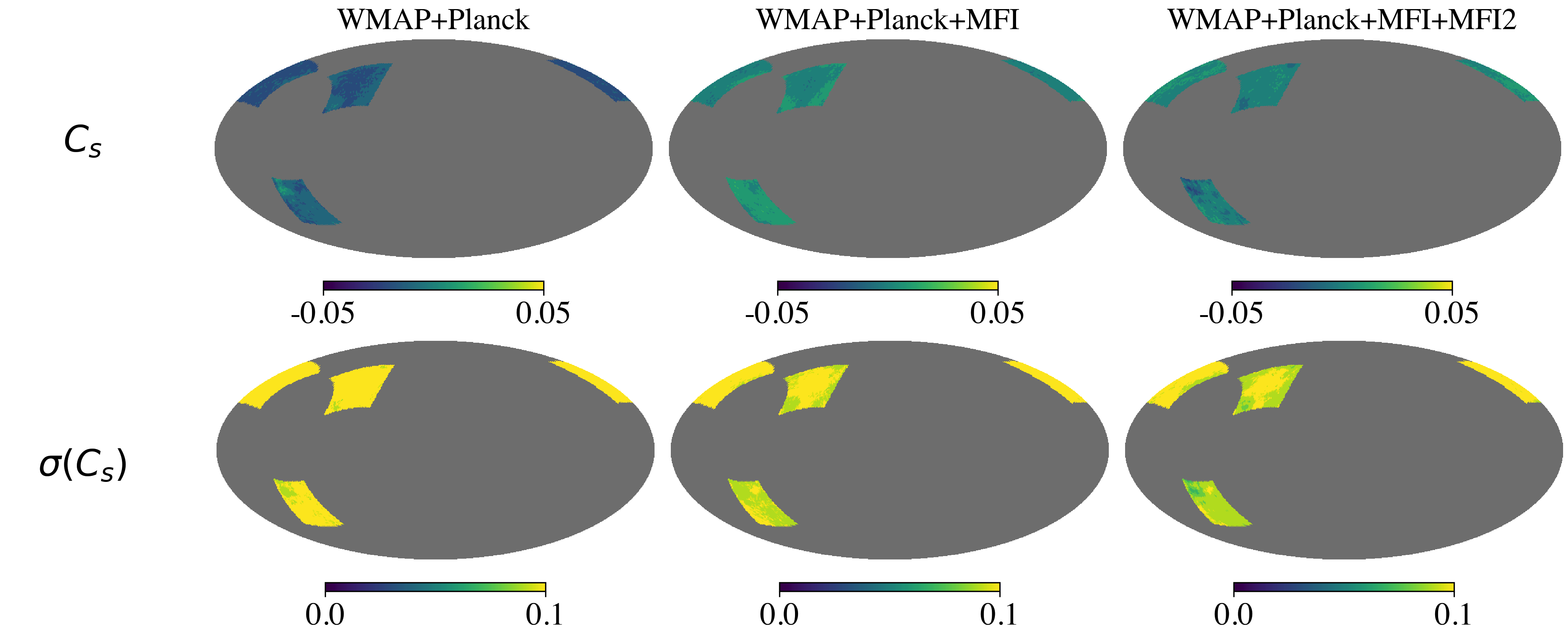}
    \caption{Maps of $C_s$ in the cosmological fields assuming a curved synchrotron model. Columns show the three datasets: WMAP+\textit{Planck} (left), WMAP+\textit{Planck}+MFI (centre), and WMAP+\textit{Planck}+MFI+MFI2 (right). Rows correspond to the recovered median $C_s$ (top) and its associated uncertainty (bottom). Grey areas indicate the masked pixels.}
    \label{fig:map_Cs_curv_cf}
\end{figure*}

Forecasts of the synchrotron curvature $C_s$ in the QUIJOTE deep cosmological fields are presented for a curved synchrotron model in Figure \ref{fig:map_Cs_curv_cf}.

Constraining the synchrotron curvature $C_s$ in the cosmological fields proves to be extremely challenging. Considering WMAP+\textit{Planck} alone, the mean SNR remains very low $\sim$0.15, with uncertainties on $C_s$ $\sim$0.1. Adding the deep-fields MFI and MFI2 channels produces only marginal improvement: the mean SNR increases slightly, while the uncertainties decrease only marginally to $\sim$0.094, with just a few isolated pixels reaching moderately higher values. 

This limited ability to constrain $C_s$ is a direct consequence of the intrinsically faint synchrotron signal of these regions. Based on the median uncertainties, a global $3 \sigma$ detection on a single pixel would require $|C_s| \geq 0.3$ for both WMAP+\textit{Planck} alone and with MFI+MFI2, values well above the expected curvature \citep{2012ApJ...753..110K}. At the forecasted sensitivities, even deep low-frequency observations within the frequency range considered here (11-19 GHz) are insufficient to detect subtle spectral curvature in these faint regions on a pixel-by-pixel basis. Detecting synchrotron curvature in such regions would therefore require substantially improved sensitivity and/or a significantly wider frequency lever arm extending to lower frequencies.

\section{Conclusions} \label{sec:conclusions}

In this paper, we have presented parametric component separation forecasts for the QUIJOTE–MFI2 instrument, aimed at quantifying its ability to improve the estimation of polarised synchrotron spectral parameters. Forecasts were carried out on maps at $1^\circ$ FWHM resolution and $N_{\rm side}=64$ pixelisation, and combining with existing public satellite data (WMAP and \textit{Planck}) as well as with the former QUIJOTE-MFI instrument, considering both the wide survey and the deep cosmological field observing modes. Two synchrotron models were considered: a pure power–law and a curved spectrum (power-law with curvature). The main conclusions are:

\begin{enumerate}
    \item WMAP+\textit{Planck} alone generally cannot constrain the synchrotron parameters on a pixel by pixel basis. The inclusion of QUIJOTE-MFI+MFI2, by contrast, yields statistically unbiased and significantly more precise measurements, with improvement factors of up to $\sim$10 for $\beta_s$, $\sim$5 for $C_s$, and $\sim$43 for the synchrotron amplitudes in bright regions of the sky. Low SNR regions benefit modestly from MFI2, with $\beta_s$ and $C_s$ improvements of up to $\sim$1.8 and $\sim$1.2, respectively.
    
    \item The addition of QUIJOTE-MFI2 data reduces uncertainties in the synchrotron spectral index $\beta_s$ by up to a factor of $\sim$7 in high-SNR regions, substantially outperforming WMAP+\textit{Planck}. In the Galactic plane, median uncertainties decrease from 0.23 (power-law) or $\sim$0.5 (curved) with WMAP+\textit{Planck} to 0.06 (power-law) or $\sim$0.08 (curved) with the inclusion of MFI+MFI2. In regions of intrinsically low signal, MFI+MFI2 reduces the uncertainty on $\beta_s$ by up to $\sim$60\%.

    \item Complex synchrotron models, such as those including spectral curvature, introduce strong degeneracies between $\beta_s$ and $C_s$, leading WMAP+\textit{Planck} alone to produce biased estimates. In contrast, the combination of QUIJOTE–MFI and MFI2 yields statistically unbiased results and substantially mitigates the degeneracy between spectral index and curvature, although it is not fully removed.

    \item In the faint, high-latitude regions, critical for cosmology studies, WMAP+\textit{Planck} alone cannot constrain the synchrotron spectral index. Only the deep QUIJOTE-MFI2 cosmological fields enable well-constrained measurements of $\beta_s$ in these low-brightness areas.

    \item Forecasted QUIJOTE-MFI2 data, in combination with WMAP and \textit{Planck}, do not provide sufficient sensitivity to detect a synchrotron curvature of $C_s=-0.052$. We find that a $2\sigma$ detection would require $|C_s|\gtrsim0.14$ in the brightest regions of the Galactic plane.

    \item In low-brightness regions of the sky, combining QUIJOTE-MFI2 with WMAP and \textit{Planck} reduces the median synchrotron residual at 100\,GHz by a factor of six, to $\sim$ 0.033 \textmu K$_{\mathrm{CMB}}$, improving the precision with which the CMB polarisation can be measured near the foreground minimum.

\end{enumerate}

Detecting synchrotron curvature on a pixel-by-pixel basis is limited not only by sensitivity but also by the frequency leverage of the current data. Meaningful constraints require additional low-frequency polarisation measurements with higher sensitivity. In particular, surveys such as the European Low Frequency Survey (ELFS, 5--20\,GHz) proposed by \cite{2024SPIE13102E..25M, ELFS2025} would help constrain the curvature by providing a wider frequency separation and thus increasing leverage and sensitivity to spectral curvature. However, other effects, such as Faraday rotation and Faraday depolarisation, may also become relevant, and the modelling presented in this work will need to be extended. 

Looking ahead, upcoming CMB surveys will greatly benefit from combining their data with deep low-frequency measurements, as these are essential for accurate synchrotron characterisation. Currently, thermal dust remains the dominant source of uncertainty over most of the sky at CMB frequencies, meaning that QUIJOTE-MFI2 and other low-frequency surveys will play a crucial role once higher-frequency dust measurements improve. In the longer term, as experiments like the Simons Observatory \citep{2019JCAP...02..056A} and LiteBIRD \citep{2023PTEP.2023d2F01L} achieve improved sensitivity and more precise dust characterisation, low-frequency synchrotron data from instruments such as QUIJOTE-MFI2 will provide valuable information for validating component separation results.

\begin{acknowledgements}
We acknowledge financial support from the Spanish MCIN/AEI/10.13039/501100011033, project references PID2020-120514GB-I00 and PID2023-151567NB-I00, and from the Horizon Europe research and innovation program under GA 101135036 (RadioForegroundsPlus). This research made use of computing time available on the high-performance computing systems at the Instituto de Astrofisica de Canarias. The authors thankfully acknowledges the technical expertise and assistance provided by the Spanish Supercomputing Network (Red Espanola de Supercomputacion), as well as the computer resources used: the DEIMOS/DIVA Supercomputer, located at the Instituto de Astrofisica de Canarias. This paper made use of the IAC HTCondor facility
(https://research.cs.wisc.edu/htcondor/), partly financed by the Ministry of Economy and Competitiveness with FEDER funds, code IACA13-3E-2493.
\end{acknowledgements}

\bibliographystyle{aa}  
\bibliography{references} 

\clearpage
\onecolumn
\begin{appendix}
\section{Surveys and frequency channels}
\label{sec:append_surveys}

\begin{table}[h!]
    \centering
     \caption{Survey channels used in this work.}
    \label{tab:surveys}
    \resizebox{\textwidth}{!}{
    \begin{tabular}{lcccccc}
        \noalign{\vskip 0.5ex}\hline\hline\noalign{\vskip 0.5ex}
        Channel & $\nu_0$ (GHz) & $\sigma_Q$ (\textmu K\,deg) & $\sigma_U$ (\textmu K\,deg) & $\theta_b$ (arcmin) & $f_\mathrm{sky}$ & Observation mode \\
        \noalign{\vskip 0.5ex}\hline\noalign{\vskip 0.5ex}
        WMAP K  & 23  & 6.9  & 7.3  & 52.8 & 1 &  \\
        WMAP Ka & 33  & 6.9  & 7.4  & 39.6 & 1 &  \\
        WMAP Q  & 41  & 6.6  & 7.1  & 30.6 & 1 & Full Sky\tablefootmark{a} \\
        WMAP V  & 61  & 7.9  & 8.5  & 21.0 & 1 & \\
        WMAP W  & 94  & 9.5  & 10.2 & 13.2 & 1 & \\
        \noalign{\vskip 0.5ex}\hline\noalign{\vskip 0.5ex}
        \textit{Planck} 30  & 28.4  & 4.5  & 4.5  & 33   & 1 &  \\
        \textit{Planck} 44  & 44.1  & 5.6  & 5.3  & 27.9 & 1 &  \\
        \textit{Planck} 70  & 70.4  & 4.6  & 4.6  & 13.1 & 1 &  \\
        \textit{Planck} 100 & 100   & 1.9  & 1.9  & 9.7  & 1 & Full Sky\tablefootmark{b}  \\
        \textit{Planck} 143 & 143   & 1.2  & 1.2  & 7.2  & 1 &  \\
        \textit{Planck} 217 & 217   & 1.8  & 1.8  & 4.9  & 1 &  \\
        \textit{Planck} 353 & 353   & 7.2  & 7.4  & 4.9  & 1 &  \\
        \noalign{\vskip 0.5ex}\hline\noalign{\vskip 0.5ex}
        QUIJOTE-MFI & 11.1 & 42.2 & 42.2 & 55.4 & 0.4  &  \\
        QUIJOTE-MFI & 12.9 & 37.9 & 37.9 & 55.8 & 0.4  &  \\
        QUIJOTE-MFI & 16.8 & 35.8 & 35.8 & 39.0 & 0.4  & WS\tablefootmark{c} \\
        QUIJOTE-MFI & 18.8 & 38.2 & 38.2 & 40.3 & 0.4  & \\
        \noalign{\vskip 0.5ex}\hline\noalign{\vskip 0.5ex}
        QUIJOTE-MFI & 11.1 & 24.7 & 24.7 & 55.4 & 0.09 &  \\
        QUIJOTE-MFI & 12.9 & 22.2 & 22.2 & 55.8 & 0.09 & \\
        QUIJOTE-MFI & 16.8 & 20.9 & 20.9 & 39.0 & 0.09 & WS + CF\tablefootmark{d} \\
        QUIJOTE-MFI & 18.8 & 22.3 & 22.3 & 40.3 & 0.09 &  \\
        \noalign{\vskip 0.5ex}\hline\noalign{\vskip 0.5ex}
        QUIJOTE-MFI2 & 10.35 & 15.9 & 15.9 & 55.4 & 0.4  &  \\
        QUIJOTE-MFI2 & 13.70  & 9.4  & 9.4  & 55.8 & 0.4  &  \\
        QUIJOTE-MFI2 & 16.4  & 11.0 & 11.0 & 39.0 & 0.4  & WS\tablefootmark{e} \\
        QUIJOTE-MFI2 & 18.75 & 10.3 & 10.3 & 40.3 & 0.4  &  \\
        \noalign{\vskip 0.5ex}\hline\noalign{\vskip 0.5ex}
        QUIJOTE-MFI2 & 10.35 & 6.2 & 6.2 & 55.4 & 0.09 &  \\
        QUIJOTE-MFI2 & 13.70  & 3.7 & 3.7 & 55.8 & 0.09 & \\
        QUIJOTE-MFI2 & 16.4  & 4.3 & 4.3 & 39.0 & 0.09 & WS + CF\tablefootmark{e}  \\
        QUIJOTE-MFI2 & 18.75 & 4.0 & 4.0 & 40.3 & 0.09 & \\
        \noalign{\vskip 0.5ex}\hline\noalign{\vskip 0.5ex}
    \end{tabular}
    }
    \tablefoot{
    Centre frequencies, sensitivities in the $Q$ and $U$ Stokes parameters, beam sizes, sky fraction and observation mode for the wide survey (WS) and cosmological fields (CF) considered in this work.\\
    \tablefoottext{a}{\citet{2013ApJS..208...20B}.}
    \tablefoottext{b}{\citet{2020AA...641A...1P}.}
    \tablefoottext{c}{\citet{mfiwidesurvey}.}
    \tablefoottext{d}{Data taken; paper in preparation.}
    \tablefoottext{e}{Forecast values (this paper).}}
\end{table}

\clearpage
\section{Jeffreys priors}
\label{sec:append_jeffreys}

\begin{table}[h!]
    \centering
    \caption{Independent Jeffreys priors for each model parameter.}
    \label{tab:jeffreys_priors}
    \begin{tabular}{l c c}
        \toprule
        \toprule
        $\theta_i$ & $\pi_{\mathrm{J}} (\theta_i)$ & Range \\
        \midrule
$A_{\mathrm{Q,s}}$ &  $\propto \sqrt{\displaystyle \sum_{j=1}^{m} \left( \frac{1}{\sigma_{Q,j}} \frac{\bar{Q}_{\mathrm{s},j}}{A_{Q,\mathrm{s}}} \right)^2}$ & [-7, 7]  $\mathrm{mK}_{\mathrm{RJ}}$ \\

$A_{\mathrm{U,s}}$ & $\propto \sqrt{\displaystyle \sum_{j=1}^{m} \left( \frac{1}{\sigma_{U,j}} \frac{\bar{U}_{\mathrm{s},j}}{A_{U,\mathrm{s}}} \right)^2}$ & [-2, 2]  $\mathrm{mK}_{\mathrm{RJ}}$ \\

$\beta_{\mathrm{s}}$ & $\propto \sqrt{\displaystyle \sum_{j=1}^{m} \log^2\left( \frac{v_j}{v_{0,s}} \right) \left( \frac{\bar{Q}_{s,j}^2}{\sigma_{Q,j}^2} + \frac{\bar{U}_{s,j}^2}{\sigma_{U,j}^2} \right)}$ & [-4, -2] \\

$C_{\mathrm{s}}$ & $\propto \sqrt{\displaystyle \sum_{j=1}^{m} \log^4\left( \frac{v_j}{v_{0,s}} \right) \left( \frac{\bar{Q}_{s,j}^2}{\sigma_{Q,j}^2} + \frac{\bar{U}_{s,j}^2}{\sigma_{U,j}^2} \right)}$ & [-0.5, 0.5] \\

$A_{\mathrm{Q,d}}$ & $\propto \sqrt{\displaystyle \sum_{j=1}^{m} \left( \frac{1}{\sigma_{Q,j}} \frac{\bar{Q}_{\mathrm{d},j}}{A_{Q,\mathrm{d}}} \right)^2}$ & [-130, 130]  $\mu\mathrm{K}_{\mathrm{RJ}}$ \\

$A_{\mathrm{U,d}}$ & $\propto \sqrt{\displaystyle \sum_{j=1}^{m} \left( \frac{1}{\sigma_{U,j}} \frac{\bar{U}_{\mathrm{d},j}}{A_{U,\mathrm{d}}} \right)^2}$ & [-100, 100]  $\mu\mathrm{K}_{\mathrm{RJ}}$ \\

$\beta_{\mathrm{d}}$ & $\propto \sqrt{\displaystyle \sum_{j=1}^{m} \log^2\left( \frac{v_j}{v_{0,d}} \right) \left( \frac{\bar{Q}_{d,j}^2}{\sigma_{Q,j}^2} + \frac{\bar{U}_{d,j}^2}{\sigma_{U,j}^2} \right)}$ & [1, 2] \\

$T_{\mathrm{d}}$ & $\propto \sqrt{\displaystyle \sum_{j=1}^{m} \frac{1}{T_d^4} \left( \frac{v_{j}}{1 - e^{-\gamma \nu }} - \frac{v_0}{1 - e^{-\gamma \nu_0 }} \right)^2 \left( \frac{\bar{Q}_{d,j}^2}{\sigma_{Q,j}^2} + \frac{\bar{U}_{d,j}^2}{\sigma_{U,j}^2} \right)}$ & [12, 30] $\mathrm{K}$ \\

$A_{\mathrm{Q,cmb}}$ & $\propto \sqrt{\displaystyle \sum_{j=1}^{m} \left( \frac{1}{\sigma_{Q,j}} \frac{\bar{Q}_{\mathrm{cmb},j}}{A_{Q,\mathrm{cmb}}} \right)^2}$ & [-2, 2]  $\mu\mathrm{K}$ \\

$A_{\mathrm{U,cmb}}$ & $\propto \sqrt{\displaystyle \sum_{j=1}^{m} \left( \frac{1}{\sigma_{U,j}} \frac{\bar{U}_{\mathrm{cmb},j}}{A_{U,\mathrm{cmb}}} \right)^2}$ & [-2, 2]  $\mu\mathrm{K}$ \\
    \bottomrule
    \end{tabular}
    \tablefoot{The summation is taken over all $m$ data points, where $\bar{Q}_{x,j}$, $\bar{U}_{x,j}$ are the theoretical prediction for the $Q$ and $U$ Stokes parameters respectively, corresponding to the model x = \{cmb, s, d\}. The upper and lower bounds on the parameters were informed by the true values extracted from the \texttt{PySM} maps. }
\end{table}

\end{appendix}

\end{document}